# Intertwined bulk photocarrier and interfacial barrier dynamics in van der Waals point-contact Schottky junctions

Munenori Yokota[1,6], Hiroyuki Mogi[2,6], Yutaka Mera[3], Katsuya Iwaya[1*], Taketoshi Minato[4,5], Shoji Yoshida[2], Osamu Takeuchi[2], Tatsuo Nakagawa[1], and Hidemi Shigekawa[2*]

[1] UNISOKU Co., Ltd., Osaka 573-0131, Japan.

[2] Faculty of Pure and Applied Sciences, University of Tsukuba, Ibaraki 305-8573, Japan.

[3] Department of Fundamental Biosciences (Physics), Shiga University of Medical Science, Shiga 520-2192, Japan.

[4] Institute for Molecular Science (IMS), National Institutes of Natural Sciences, Aichi 444-8585, Japan.

[5] Core for Spin Life Sciences, Okazaki Collaborative Platform, National Institutes of Natural Sciences, Aichi 444-8585, Japan.

[6] These authors contributed equally: Munenori Yokota, Hiroyuki Mogi.

Corresponding author: Katsuya Iwaya (katsuya_iwaya@unisoku.co.jp)
Hidemi Shigekawa (hidemi@ims.tsukuba.ac.jp)

**Abstract**

Schottky junctions based on transition-metal dichalcogenides underpin next-generation optoelectronic devices, yet the nanoscale carrier dynamics that ultimately govern their performance remain largely inaccessible. Here, we employ optical pump–probe time-resolved atomic force microscopy (OPP-TR-AFM) to investigate the spatiotemporal response of a $WSe_2$/PtIr van der Waals Schottky junction. Under backside illumination, photocarriers generated in bulk $WSe_2$ produce a local junction response through coupled bulk carrier transport and interfacial barrier modulation. We show that the time-resolved current is consistently explained by the competition between long-lived hole diffusion governed by the occupancy of bulk trap states and thermionic emission current modulated by the surface photovoltage. This mechanism is supported by transient absorption spectroscopy, light-modulated current–voltage measurements, and numerical simulations. Furthermore, spatial mapping reveals pronounced nanoscale variations in the coupled dynamics of carrier transport and interfacial barrier modulation. These results demonstrate that OPP-TR-AFM provides a new framework for resolving the coupled behavior of bulk carriers and interfacial barrier dynamics in van der Waals Schottky junctions with simultaneous spatial and temporal resolution, establishing it as a powerful approach for investigating the microscopic processes that play a crucial role in the development of next-generation optoelectronic devices.

## INTRODUCTION

Transition metal dichalcogenides (TMDCs) have attracted considerable attention as promising materials for next-generation optoelectronic devices owing to their strong light–matter interactions, tunable electronic structures, and excellent carrier transport properties[1,2,3]. In particular, TMDC-metal Schottky junctions play a central role in a wide range of optoelectronic technologies, including photodetectors and energy-harvesting devices[4,5]. Recently, $WSe_2$-based Schottky junctions have demonstrated nanosecond-scale photoresponses, highlighting their potential for high-speed optoelectronic applications[6,7,8]. However, the microscopic processes governing carrier transport and photocurrent generation remain poorly understood.

A central challenge is that photocurrent response in TMDC Schottky junctions is determined not only by carrier generation and transport within the semiconductor but also by the dynamic response of the interfacial energy landscape, including band bending, surface photovoltage (SPV), and trap-mediated processes[9,10,11,12,13,14,15]. Because these processes evolve on nanometer and nanosecond scales, conventional spatially averaged techniques, such as pump–probe spectroscopy and device-level electrical measurements, have limited ability to directly resolve the local interplay among them.

To address this challenge, we employed optical pump–probe time-resolved atomic force microscopy (OPP-TR-AFM), which combines optical pump–probe excitation with local electrical measurements to achieve simultaneous nanosecond temporal resolution and nanometer spatial resolution. A local van der Waals Schottky junction was formed between a conductive AFM probe and the surface of bulk $WSe_2$[16], providing a junction geometry relevant to practical Schottky optoelectronic devices. In the backside-illumination configuration, photoexcited carriers are generated predominantly within the bulk $WSe_2$ and subsequently diffuse toward the interface. Consequently, the measured photocurrent reflects the coupled dynamics of bulk carrier transport and local interfacial barrier modulation.

To elucidate the origin of the observed response, we combined OPP-TR-AFM measurements with transient absorption spectroscopy (TAS) covering timescales from picoseconds to microseconds[17], light-modulated current–voltage (*I*-*V*) measurements[18], and numerical simulations. We demonstrate that the time-resolved current arises from the competition between long-lived hole diffusion governed by bulk trap occupancy and thermionic emission current modulated by the surface photovoltage. Furthermore, spatial mapping reveals pronounced nanoscale variations in these coupled processes. Together, these findings establish OPP-TR-AFM as a powerful approach for investigating the interplay between bulk carrier transport and interfacial barrier dynamics, providing new insight into the microscopic operation of van der Waals Schottky optoelectronic devices.

## RESULTS AND DISCUSSION

### Experimental configuration and operating principle

Figure 1a schematically illustrates the OPP-TR-AFM measurement system used in this study (see

Supplementary Fig. 1 for details). A local van der Waals Schottky junction was formed by gently bringing a PtIr-coated conductive AFM cantilever into point contact with the surface of bulk $WSe_2$, providing a local junction geometry relevant to practical Schottky optoelectronic devices. Optical excitation was performed under backside illumination through a transparent substrate (see Methods). In this illumination geometry, photoexcited carriers are generated predominantly within the bulk $WSe_2$ and subsequently transported through the crystal, with a fraction reaching the local $WSe_2$/PtIr interface. Consequently, the measured time-resolved current reflects both carrier transport within the bulk and the temporal evolution of the interfacial Schottky barrier.

Figure 1b shows the light-modulated *I–V* characteristics of the $WSe_2$/PtIr junction measured under periodic light ON/OFF modulation. This measurement enables simultaneous evaluation of the current responses under dark and illuminated conditions. The *I–V* characteristics can be divided into three bias regimes: (A) the forward-bias region (sample bias, $V_s < -0.3$ V), (B) the reverse-bias region ($-0.3 < V_s < +1.5$ V), and (C) the tunneling regime ($V_s > +1.5$ V).

Under dark conditions, the *I–V* characteristics exhibit the rectifying behavior characteristic of a Schottky junction, consistent with the $WSe_2$ surface behaving effectively as an *n*-type semiconductor. Analysis of the forward-bias characteristics in region (A) based on the thermionic emission model[19,20,21] yielded an effective Schottky barrier height of approximately 0.65 eV (Supplementary Fig. 2). In the reverse-bias region (B), the thermionic emission current $I_{th}$ over the Schottky barrier is the dominant current component (Fig. 1c). At higher bias, in the tunneling regime (C), the tunneling current $I_{tunnel}$ becomes dominant, suggesting that the depletion region near the $WSe_2$ surface is relatively narrow.

Under illumination, the onset voltage for tunneling conduction $V_{tunnel}$ shifts systematically toward lower sample bias with increasing excitation fluence (Fig. 1b). This behavior indicates that optical excitation induces an SPV, thereby modifying the electrostatic potential near the interface[22,23]. We attribute this shift to the transient accumulation of photoexcited holes in localized states near the interface, which partially relaxes the band bending and consequently reduces the sample bias required to initiate tunneling conduction.

Based on these observations, the carrier-transport model considered in this study is illustrated in Fig. 1d. Photoexcited holes generated in the bulk are transported toward the interface through trap-mediated diffusion and contribute to the hole current $I_{hole}$. Meanwhile, photoexcited charges accumulated in localized states near the interface partially relax the band bending, thereby modulating the Schottky barrier and changing $I_{th}$. Here, the interfacial localized states refer to localized electronic states that may form at the metal–semiconductor interface, such as those arising from local structural disorder at the $WSe_2$/PtIr contact or metal-induced gap states[24].

As shown below, the OPP-TR-AFM measurements exhibit characteristic bias- and excitation-fluence-dependent behaviors, particularly in bias region (B). To elucidate these behaviors, we discuss the origins of the delay-time, bias, and excitation-fluence dependences of the time-resolved current in terms of the contributions from $I_{hole}$ and $I_{th}$ modulated by the SPV.

**OPP-TR-AFM measurements**

Next, time-resolved current signals were measured using pump–probe pulse pairs in combination with the delay-time modulation method[25] (see Methods and Supplementary Fig. S1). The delay-time modulation method enables highly sensitive detection of the temporal response of the Schottky junction while effectively suppressing thermal background signals. In addition, no significant change in the time-resolved current was observed when the tip contact force was varied, confirming that the time-resolved response of the van der Waals Schottky junction is measured reproducibly under the present experimental conditions (Supplementary Fig. S3).

Representative OPP-TR-AFM measurements are shown in Figs. 2a,b. The time-resolved current exhibits pronounced dependence on both sample bias and excitation fluence. Figure 2a shows the bias dependence measured at an excitation fluence of 3.4 μJ cm$^{-2}$. At low bias ($V_s$ = +0.3 V), only a negative response is observed. As the bias increases, a positive component emerges, and at $V_s$ = +0.6 V, both positive and negative components coexist. At higher biases (+1.0 and +1.5 V), the positive component becomes dominant. To further characterize this behavior, we also analyzed the bias dependences of the illuminated DC current (average current), measured without delay-time modulation, and the time-resolved current at zero delay, $\Delta I_L(t_d = 0)$ (Supplementary Fig. S4).

Figure 2b shows the excitation-fluence dependence measured at $V_s$ = +0.3 V. At the maximum excitation fluence (100%), a large negative response is observed. As the excitation fluence decreases, however, the negative component is rapidly suppressed and a positive component appears instead. The positive component reaches its maximum at 50% excitation fluence and decreases again at 10%. These observations indicate that the measured time-resolved current originates from the competition between two processes with distinct excitation-fluence dependences, and that their relative contributions are strongly governed by the photoexcited carrier density. Furthermore, the excitation-fluence dependence of the photocurrent obtained from the light-modulated *I–V* measurements was also analyzed (Supplementary Fig. S5), providing complementary information on the excitation-density dependence of the photoresponse.

The amplitudes and lifetimes obtained by exponential fitting are summarized in Fig. 2c,d, respectively. As shown in Fig. 2c, the amplitude of the positive component increases with sample bias and becomes dominant at higher bias, particularly under the 100% excitation condition. In contrast, the negative component is observed only at the maximum excitation fluence (100%) in the low-bias region and gradually decreases with increasing bias.

As shown in Fig. 2d, the lifetimes also exhibit distinct behaviors for the positive and negative components. The lifetime of the negative component decreases from approximately 40 ns at +0.3 V to approximately 20 ns at +0.6 V under the 100% excitation condition. In contrast, the positive component exhibits a lifetime of approximately 50–90 ns, and shows weak bias dependence particularly under the 10% and 50% excitation conditions.

These results indicate that the time-resolved current consists of two components with opposite polarities, each exhibiting distinct bias and excitation-fluence dependences. In the following section, we discuss the physical origins of these two time-resolved current components and how their competition determines the observed temporal response.

**Nanosecond carrier dynamics in bulk $WSe_2$**

To characterize the optical response of bulk $WSe_2$, TAS covering the picosecond-to-microsecond time range was performed using the Randomly Interleaved Pulse Train (RIPT) method[17] (see Methods). Figure 3a shows the steady-state absorption spectrum of bulk $WSe_2$. The absorption features labeled A and B correspond to the A (~1.65 eV) and B (~2.1 eV) excitonic transitions at the K point, respectively, while the broad C band originates from higher-energy transitions in the band-nesting region[26] (Supplementary Fig. 6). The 488 nm pump pulse (2.54 eV) used in this study predominantly excites carriers through the C-band transition. Because the exciton binding energy in bulk $WSe_2$ is approximately 30 meV, comparable to the thermal energy at room temperature, photoexcited excitons readily dissociate into free carriers[27,28]. Consequently, the nanosecond dynamics discussed in this study are expected to be governed predominantly by free carriers.

Representative differential transmission $\Delta T/T$ spectra are shown in Fig. 3b. Positive transient features ($A^+$ and $B^+$) are observed near the A and B excitonic transitions, accompanied by corresponding negative sidebands ($A^-$ and $B^-$) on their lower-energy sides. Similar spectral features have previously been reported in transient absorption studies of bulk TMDCs, primarily over the sub-picosecond to nanosecond time range[29]. By employing the RIPT method, the present measurements extend these observations to the previously unexplored nanosecond-to-microsecond regime. The $A^+$ and $B^+$ features, together with their corresponding sidebands, evolve continuously with delay time. Such spectral evolution has been attributed to energy shifts and linewidth changes of the excitonic absorption arising from interactions between photoexcited free carriers and excitonic resonances[30].

At longer delay times, however, the transient spectra may also contain contributions from photoinduced heating. To evaluate this effect, we analyzed the energy shift and linewidth broadening of the excitonic resonance based on the transient absorption spectra (Supplementary Fig. 7). The analysis suggests that nonthermal interactions between the remaining photoexcited carriers and the excitonic resonances play a dominant role in the transient spectral evolution under the low-excitation conditions corresponding most closely to those employed in the OPP-TR-AFM measurements. Accordingly, the relaxation dynamics obtained by TAS were used as a reference for the temporal evolution of bulk photocarriers when compared with the time-resolved current measured by OPP-TR-

AFM.

Figure 3c shows the delay-time dependence of $\Delta T/T$ at 760 nm ($A^+$). The transient signal persists over several tens to hundreds of nanoseconds. To quantify the observed dynamics, the data were fitted using a phenomenological triple-exponential function. The decay times extracted from the fitting are summarized in Fig. 3d for each excitation fluence. At the lowest excitation fluence of 3 μJ $cm^{-2}$, comparable to that used in the OPP-TR-AFM measurements, the extracted decay times were $\tau_1 \approx 0.9$ ns, $\tau_2 \approx 3.3$ ns, and $\tau_3 \approx 75.9 \pm 30$ ns (Supplementary Fig. 6). For comparison, the lifetimes obtained from the OPP-TR-AFM measurements are also plotted as open symbols. The lifetime of the positive time-resolved response observed by OPP-TR-AFM falls within the same tens-of-nanoseconds time range as the longest-lived component obtained from TAS. This correspondence in timescales supports the interpretation that the positive OPP-TR-AFM response is closely associated with long-lived photocarriers generated in the bulk $WSe_2$ and subsequently diffusing toward the interface. It should also be noted that the $\Delta T/T$ signal is nearly zero at the excitation wavelength of 488 nm used for the OPP-TR-AFM measurements, indicating that the OPP-TR-AFM signal cannot be explained simply by changes in optical absorption. Taken together, these results support the interpretation that the OPP-TR-AFM signal reflects the coupled dynamics of long-lived carrier transport in the bulk and the response of the local Schottky junction.

**Origin of the positive and negative time-resolved current responses**

To understand the origin of the positive and negative time-resolved current responses observed in the OPP-TR-AFM measurements, we constructed the carrier transport model summarized in Fig. 4. We propose that the time-resolved response in the reverse-bias regime can be understood as the competition between the hole diffusion current $I_{\mathrm{hole}}$, arising from photogenerated holes diffusing through the bulk, and the thermionic emission current $I_{\mathrm{th}}$, modulated by photoinduced changes in the interfacial potential.

We first consider $I_{\mathrm{hole}}$. Based on the light-modulated $I$–$V$ measurements, the $WSe_2$ surface is treated as an effectively *n*-type semiconductor with a dark electron density of $N_{\mathrm{dark}}$. Under backside illumination, photoexcited carriers are generated predominantly within the bulk $WSe_2$ region and subsequently diffuse toward the $WSe_2$/PtIr interface. As shown in Fig. 3d, bulk $WSe_2$ exhibits a long-lived optical response extending over several tens of nanoseconds. To describe the relaxation dynamics of bulk photocarriers, we employ the Shockley–Read–Hall (SRH) recombination model[31,32].

As illustrated in Figs. 4a,b, the effective hole lifetime is strongly influenced by the occupancy of trap states in the bulk. When the photocarrier densities ($\Delta N_{\mathrm{e}}$, $\Delta N_{\mathrm{p}}$) are lower than $N_{\mathrm{dark}}$ (Fig. 4a), most trap states remain occupied by electrons, allowing trap-assisted hole recombination to proceed efficiently and resulting in a relatively short effective hole lifetime. In contrast, when $\Delta N_{\mathrm{e}}$ and $\Delta N_{\mathrm{p}}$ exceed $N_{\mathrm{dark}}$ (Fig. 4b), the electron occupancy of the trap states decreases, reducing the probability of hole capture and thereby increasing the effective hole lifetime. Consequently, a larger fraction of the photogenerated holes reaches the interface and contributes to $I_{hole}$. To further support this picture, we calculated the hole lifetime based on the SRH model by explicitly considering the occupancy of trap states[33] (Supplementary Fig. S8). The calculated effective hole lifetime increases with increasing photocarrier density ($\Delta N_p = \Delta N_e$), consistent with the proposed SRH-based description of the hole diffusion current.

The corresponding pump–probe response is illustrated in Fig. 4c. At short delay times, the pump pulse generates relatively high photocarrier densities ($\Delta N_e$, $\Delta N_p$), corresponding to the situation shown in Fig. 4b, resulting in a large hole diffusion current induced by the probe pulse. As the delay time increases, the photocarrier density gradually decreases, and the system evolves toward the condition shown in Fig. 4a. Consequently, the effective hole lifetime becomes shorter, and the hole diffusion current decays with time. This behavior appears experimentally as the positive time-resolved component.

We next consider $I_{th}$, arising from photoinduced modulation of the effective Schottky barrier. In the dark, the reverse-bias current is dominated by thermionic emission across the $WSe_2$/PtIr Schottky junction (Fig. 4d). Upon photoexcitation, photogenerated charges are transiently accumulated in interfacial localized states associated with the local $WSe_2$/PtIr contact. The accumulated charges partially screen the built-in electric field, thereby generating an SPV (Fig. 4e). This modifies the effective Schottky barrier and thereby changes $I_{th}$. At sufficiently high excitation densities, the accumulation of interfacial charge approaches saturation (Fig. 4f).

The corresponding pump–probe response is illustrated in Fig. 4g. At short delay times, a large population of pump-generated carriers remains near the interface, resulting in a high occupancy of the interfacial states. Consequently, only a small amount of additional interfacial charge is accumulated by the probe pulse, leading to a small photoinduced modulation of $I_{th}$. As the delay time increases, the pump-generated carriers gradually relax, allowing the probe pulse to induce a larger accumulation of interfacial charge and, consequently, a larger photoinduced modulation of $I_{th}$. Therefore, the time-resolved current measured at short delay times is smaller than that at the reference delay time $t_{max}$, giving rise to a negative $\Delta I_{L}(t_{D})$. As the delay time increases, $I_{L}(t_{d} = t_{D})$ gradually approaches the reference value $I_{L}(t_{d} = t_{max})$, and $\Delta I_{L}(t_{D})$ correspondingly recovers toward zero.

Numerical simulations based on these two competing processes successfully reproduce the principal experimental observations (see the Supplementary Information for details). Figure 4h shows the excitation-fluence dependence at $V_s = +0.3$ V. At high excitation fluences, the negative time-resolved response dominates. As the excitation fluence decreases, however, the negative component is progressively suppressed, while the positive component emerges, in good agreement with the experimental results shown in Fig. 2b.

Figure 4i shows the bias dependence at a fixed excitation fluence. This behavior can be understood in terms of the relaxation dynamics of the interfacial charge population. In the model, increasing the reverse bias shortens the relaxation time of the photoinduced interfacial potential (Eq. S10 in the Supplementary Information), leading to more rapid recovery of the effective Schottky barrier. Consequently, the negative time-resolved response originating from photoinduced modulation of $I_{th}$ is reduced, whereas the relative contribution of the positive response arising from $I_{hole}$ increases. As a result, the simulated response gradually changes from a negative signal at low bias to a positive response at high bias, reproducing the experimentally observed behavior.

In this model, $I_{hole}$ is the primary origin of the positive time-resolved response, whereas $I_{th}$ is primarily responsible for the negative response. However, because $I_{th}$ is modulated by the SPV dynamics associated with hole trapping and detrapping near the interface, it also contains a small component originating from the hole diffusion process (Supplementary Fig. S9).

These results demonstrate that the characteristic bias and excitation-fluence dependences observed by OPP-TR-AFM arise from the competition between $I_{hole}$, driven by long-lived bulk photocarriers, and $I_{th}$, modulated by the interfacial potential. The present model provides a physical framework linking bulk photocarrier transport with local interfacial barrier dynamics in van der Waals Schottky junctions.

**Spatial mapping of the time-resolved current**

A key advantage of OPP-TR-AFM is its ability to simultaneously achieve nanometer-scale spatial resolution and nanosecond temporal resolution. To demonstrate this capability, we performed spatial mapping of the time-resolved current over a 4 μm × 4 μm region on the $WSe_2$ surface.

Figures 5a and 5b show the contact-mode AFM image and the simultaneously acquired DC current map at a sample bias of $V_s = +0.6$ V, respectively. While the surface topography is relatively uniform over the entire measurement area, the DC current exhibits pronounced spatial inhomogeneity, indicating that the local junction characteristics vary across the sample.

Figures 5c–f present spatial maps of the time-resolved current acquired at delay times of 0, 30, 60, and 120 ns. At $t_d = 0$ ns, both positive and negative time-resolved responses coexist within the same field of view. This observation indicates that the relative contributions of $I_{hole}$ and the photoinduced modulation of $I_{th}$ vary locally depending on the junction characteristics.

To investigate the origin of this spatial inhomogeneity, light-modulated *I–V* measurements were performed at the representative locations indicated by the magenta and blue circles in Fig. 5a. The corresponding results are shown in Fig. 5g, h. The illuminated *I–V* characteristics differ substantially between the two locations. In particular, the magenta position exhibits a larger photoinduced shift of the *I–V* characteristics than the blue position, together with a shift of the forward current onset toward lower bias. These observations suggest that photoinduced Schottky barrier modulation contributes more strongly at the magenta position.

The corresponding time-resolved current responses are shown in Fig. 5i, j. At the magenta position, both positive and negative components are observed, whereas only a positive time-resolved response appears at the blue position. These results suggest that the photoinduced modulation of $I_{th}$ makes a larger contribution at the magenta position, while $I_{hole}$ dominates at the blue position. The fitting analysis (Fig. 5k) yields long-lived components with lifetimes of approximately 55–90 ns, in good agreement with the longest-lived optical response observed by TAS. This correspondence further supports the interpretation presented in the previous section that the positive time-resolved response originates from long-lived bulk photocarriers.

Importantly, the spatial inhomogeneity observed in this study reflects the coupled dynamics of bulk photocarriers and the local interfacial response. Photocarriers generated within the bulk $WSe_2$ by backside illumination subsequently diffuse toward the $WSe_2$/PtIr interface, where they interact with the local Schottky junction. Therefore, the observed spatial variations are likely to arise from multiple factors, including spatial variations in the effective Schottky barrier height, the density of interfacial localized states, local variations in bulk carrier transport associated with trap-assisted recombination. Further investigation will be required to distinguish the relative contributions of these factors. These results directly reveal that the coupling between bulk photocarrier transport and the local interfacial response varies substantially across the junction on the nanoscale.

Using OPP-TR-AFM in combination with light-modulated *I*–*V* measurements, transient absorption spectroscopy, and numerical simulations, we elucidated the mechanism underlying the photoinduced time-resolved current in a $WSe_2$/PtIr point-contact van der Waals Schottky junction. The observed response is explained by the competition between the diffusion of long-lived bulk photocarriers toward the interface and the photoinduced modulation of thermionic emission through transient changes in the effective Schottky barrier. Furthermore, spatially resolved measurements revealed pronounced nanoscale inhomogeneity in the time-resolved current response, revealing that the interplay between bulk photocarrier transport and local interfacial dynamics varies locally across the junction.

These findings demonstrate that OPP-TR-AFM provides a unique approach for investigating the interplay between bulk carrier transport and local interfacial dynamics with simultaneous nanometer-scale spatial resolution and nanosecond temporal resolution. Beyond TMDC-based optoelectronic devices, this technique is expected to be broadly applicable to a wide range of van der Waals heterostructures and low-dimensional semiconductor junctions, providing new opportunities to elucidate nanoscale carrier transport and interfacial dynamics and to guide the design and optimization of next-generation optoelectronic and energy-harvesting devices.

## METHODS

### Optical pump-probe time-resolved current

The optical pump–probe delay time is denoted by $t_{\mathrm{d}}$. During the measurement, a specific delay time $t_{\mathrm{d}} = t_{\mathrm{D}}$ is selected, and the corresponding photoinduced transient current $I_{\mathrm{L}}^{*}(t_{\mathrm{D}})$ persists until the pump–probe excited state has completely relaxed. Accordingly, the measured time-resolved signal $I_{\mathrm{L}}(t_{\mathrm{D}})$ is given by the cycle-averaged transient current over one laser repetition period $\tau_{\mathrm{R}}$,

$$I_{\mathrm{L}}(t_{\mathrm{D}}) = \frac{1}{\tau_{\mathrm{R}}}\int_{0}^{\tau_{\mathrm{R}}} I_{\mathrm{L}}^{*}(t_{\mathrm{d}} = t_{\mathrm{D}})dt \qquad (1)$$

To detect the small current variation with high sensitivity, lock-in detection was employed. Furthermore, to minimize the influence of photothermal expansion of both the tip and the sample, the delay-time modulation method was used[25]. In this approach, instead of modulating the pump or probe pulses themselves as the lock-in reference, the pump–probe delay is alternately switched between the delay of interest $t_{\mathrm{d}} = t_{\mathrm{D}}$ and a sufficiently long delay $t_{\mathrm{d}} = t_{\mathrm{max}}$. The lock-in signal is then obtained as the differential current between these two delay conditions,

$$\Delta I_{\mathrm{L}}(t_{\mathrm{D}}) = I_{\mathrm{L}}(t_{\mathrm{d}} = t_{\mathrm{D}}) - I_{\mathrm{L}}(t_{\mathrm{d}} = t_{\mathrm{max}}) \qquad (2)$$

This method enables highly sensitive extraction of the temporal response of the Schottky junction while minimizing thermal artifacts arising from laser irradiation.

### Optical system for OPP-TR-AFM

The laser pulses were precisely synchronized using electronic triggering, enabling a simple configuration that does not require a complex optical setup (OPP-NS, Unisoku)[34],[35]. In this study, we focused on nanosecond dynamics associated with the Schottky barrier and localized interfacial states. Accordingly, two nanosecond pulsed lasers (NPL49B, Thorlabs) were used as the pump and probe sources. Both lasers operated at a wavelength of 488 nm with a pulse duration of approximately 6 ns and a repetition rate of 1 MHz. The timing jitter between the two lasers relative to the electronic trigger was approximately 20 ps, resulting in an estimated overall temporal resolution of approximately 8.5 ns. The system was integrated with a commercial AFM operated in air (Dimension XR Icon, Bruker) to enable its application to a wide range of materials and devices. All measurements were performed at room temperature in air using a PtIr-coated conductive cantilever (PF-TUNA, Bruker). For the light-modulated current–voltage measurements, the illumination was periodically switched on and off using an optical chopper.

### Sample preparation

To enable backside optical excitation, a bulk $WSe_2$ single crystal (HQ Graphene) was mounted on a conductive ITO-coated glass substrate using a transparent insulating epoxy adhesive (FH5313PK, Thorlabs) (Fig. 1a). The $WSe_2$ surface was prepared by mechanical exfoliation, and its atomically flat surface was confirmed by STM (Supplementary Fig. 1). To establish an electrical connection between the $WSe_2$ crystal and the ITO electrode, conductive silver epoxy (EPO-TEK H20E) was applied to the edge of the crystal.

During the measurements, the sample was placed on the glass plate of the AFM stage and electrically contacted through a magnetic contact pin, which also served as the sample-bias electrode (Fig. 1a). In this configuration, the illuminated region of the $WSe_2$ crystal was electrically isolated from the ITO substrate, and the $WSe_2$ crystal was connected to the ITO electrode only through the conductive silver epoxy located outside the illuminated region.

The measurement geometry contained two metal–$WSe_2$ junctions: the PtIr-coated AFM cantilever–$WSe_2$ junction and the silver-epoxy–$WSe_2$ junction. However, as demonstrated in previous studies[19], the measured current is dominated by the local junction directly beneath the AFM tip, owing to the orders-of-magnitude smaller contact area of the PtIr/$WSe_2$ junction compared with the silver-epoxy/$WSe_2$ contact.

**Picosecond-to-microsecond transient absorption spectroscopy (ps-TAS)**

In conventional transient absorption spectroscopy, the time window from approximately 1 ns to several tens of nanoseconds has been a "blind gap," largely because it is difficult to continuously vary the pump–probe delay within this regime using standard delay-line optics. To overcome this limitation, we utilized a commercially available picosecond transient absorption system, picoTAS (Unisoku), based on the RIPT method[17]. This technique, which we previously developed, enables asynchronous and randomized sampling of pump–probe delays using a picosecond supercontinuum laser as the probe light, thereby providing access to this otherwise inaccessible temporal region. For the pump pulses, we used a 1 kHz, 25 ps mode-locked picosecond laser equipped with a tunable wavelength unit (PL2210A and PG403, EKSPLA). The pump wavelength was set to 488 nm to allow direct comparison with the OPP-TR-AFM measurements. As the probe source, we used a 20 MHz, 50-100 ps supercontinuum laser (SMHP-20.2-A, Leukos), spectrally selected by a diffraction-grating monochromator to cover the wavelength range from 450 to 900 nm. Temporal resolution of the system was better than 100 ps and was primarily determined by the pulse width of the probe light. All ps-TAS measurements were performed at room temperature under ambient conditions.

**DECLARATION STATEMENTS**

**Data Availability**

The datasets generated and/or analyzed during the current study are not publicly available because they are too large to be deposited in a public repository but are available from the corresponding author upon reasonable request.

**Acknowledgements**

The authors express their gratitude to Tadashi Ueda (IMS) and Shunji Yamamoto (Unisoku) for their technical support. This work was supported by JSPS KAKENHI (Grant Nos. 23H00264, 24H00416, and 25H00835) and by JST PRESTO (Grant No. JPMJPR22AA).

**Author Contributions**

OPP-TR-AFM experiments were performed by M.Y., K.I., and T.M., while the ps-TAS measurements were carried out by T.N. S.Y. and O.T. provided technical advice on sample preparation and experimental methodologies. The mechanism of photoinduced dynamics at the Schottky junction was primarily conceived by H.M., Y.M., M.Y., and H.S., and was further established through combined investigations of time-resolved signal simulations by Y.M. and analyses of *I–V* characteristics by H.M. Based on these experimental and theoretical results, all authors discussed the details and jointly prepared the manuscript. The project was organized and supervised by K.I. and H.S., who coordinated the overall research effort.

**Competing Interests**

The authors declare no competing financial or non-financial interests.

## FIGURE LEGENDS

**Fig. 1 OPP-TR-AFM configuration and carrier transport model of the $WSe_2$/PtIr van der Waals Schottky junction.**

(a) Schematic illustration of the experimental configuration. Optical excitation was performed under backside illumination through a transparent substrate. The illuminated region of the $WSe_2$ crystal was electrically isolated from the ITO substrate by a transparent insulating adhesive layer, while the $WSe_2$ crystal was electrically connected to the ITO substrate only through a conductive adhesive applied outside the illuminated region.

(b) Light-modulated *I–V* characteristics measured at different excitation fluences. The bias dependence is divided into three regimes: (A) the forward-bias region, (B) the reverse-bias region, and (C) the tunneling regime. The OPP-TR-AFM measurements presented in this study were performed primarily in region (B). The threshold bias for the onset of tunneling conduction is denoted by $V_{tunnel}$.

(c) Schematic energy-band diagram of the $WSe_2$/PtIr junction in the dark. In the reverse-bias region, the current is dominated by the thermionic emission current $I_{th}$ over the effective Schottky barrier. At sufficiently high bias, the tunneling current $I_{tunnel}$ also contributes to the measured current.

(d) Schematic illustration of the carrier transport processes under optical excitation. Photoexcited carriers are generated predominantly within the bulk $WSe_2$ region and subsequently diffuse toward the $WSe_2$/PtIr interface. Photoexcited holes contribute to the hole diffusion current $I_{hole}$, whereas charge accumulation in localized states near the interface modulates the effective Schottky barrier, thereby changing the thermionic emission current $I_{th}$.

**Fig. 2 Bias- and excitation-fluence dependences of the time-resolved current in the $WSe_2$/PtIr junction.**

(a) Bias-voltage dependence of the time-resolved current $\Delta I_L(t_d)$ measured at an excitation fluence of 3.4 μJ cm$^{-2}$ with a contact force of 12 nN.

(b) Excitation-fluence dependence of $\Delta I_L(t_d)$ measured at a sample bias of $V_s$ = +0.3 V with a contact force of 12 nN. An excitation fluence of 100% corresponds to 3.4 μJ cm$^{-2}$.

(c, d) Bias-voltage dependences of the amplitudes and lifetimes obtained by exponential fitting of $\Delta I_L(t_d)$ measured at excitation fluences of 10%, 50%, and 100%.

**Fig. 3 Photocarrier dynamics in bulk $WSe_2$ revealed by transient absorption spectroscopy.**

(a) Steady-state absorption spectrum of bulk $WSe_2$. The A and B peaks correspond to the excitonic transitions at the K point, while the broad C band originates from higher-energy transitions in the band-nesting region.

(b) Representative differential transmission spectra $\Delta T/T$ of bulk $WSe_2$ measured at selected delay times under 488 nm excitation (pump fluence: 300 μJ cm$^{-2}$).

(c) Delay-time dependence of $\Delta T/T$ measured at 760 nm ($A^+$) for different pump fluences. Solid lines represent fits using a phenomenological triple-exponential function.

(d) Pump-fluence dependence of the decay times extracted from the triple-exponential fitting. Open symbols indicate the lifetime of the positive component obtained from the OPP-TR-AFM measurements at 3.4 μJ cm$^{-2}$ (Fig. 2d).

**Fig. 4 Physical model and numerical simulations of the time-resolved current responses in the $WSe_2$/PtIr junction.**

(a, b) Schematic illustration of the mechanism responsible for $I_{hole}$. Changes in the occupancy of bulk trap states modify the effective hole lifetime, resulting in an increase or decrease in $I_{hole}$. Here, $\Delta N_e$ and $\Delta N_p$ denote the photoexcited electron and hole densities, respectively, and $N_{dark}$ is the equilibrium electron density in the dark.

(c) Schematic pump–probe response corresponding to $I_{hole}$, giving rise to the positive time-resolved signal.

(d–f) Schematic illustration of the photoinduced modulation mechanism of $I_{th}$. Charge accumulation near the interface gives rise to SPV, thereby modulating the effective Schottky barrier. The simulated time evolution of the photoinduced interface potential is shown in Supplementary Fig. S10.

(g) Schematic pump–probe response corresponding to the photoinduced modulation of $I_{th}$,

giving rise to the negative time-resolved signal.

(h) Simulated excitation-fluence dependence of the time-resolved current at $V_s$ = +0.3 V.

(i) Simulated bias-voltage dependence of the time-resolved current at a fixed excitation fluence.

**Fig. 5 Spatially resolved time-resolved currents and local photocurrent dynamics in the $WSe_2$/PtIr junction.**

(a) Contact-mode AFM image (4 μm × 4 μm, setpoint force: 16.8 nN).

(b) Simultaneously acquired DC current map at $V_s$ = +0.6 V. The circles indicate the locations where the light-modulated *I*–*V* characteristics and time-resolved currents were measured.

(c–f) Spatial maps of the time-resolved current acquired at delay times of $t_d$ = 0, 30, 60, and 120 ns ($V_s$ = +0.6 V).

(g, h) Light-modulated *I*–*V* characteristics measured at the blue and magenta locations, respectively (excitation fluence: 6.7 μJ $cm^{-2}$).

(i, j) Delay-time dependences of the time-resolved current measured at the magenta and blue locations, respectively ($V_s$ = +0.6 V, setpoint force: 14.4 nN).

(k) Amplitudes $A_1$, $A_2$ and lifetimes $\tau_1$, $\tau_2$ obtained by exponential fitting of the time-resolved currents shown in (i) and (j). The response at the magenta location was fitted with two components of opposite polarity, whereas that at the blue location was fitted with a single positive component.

Fig. 1

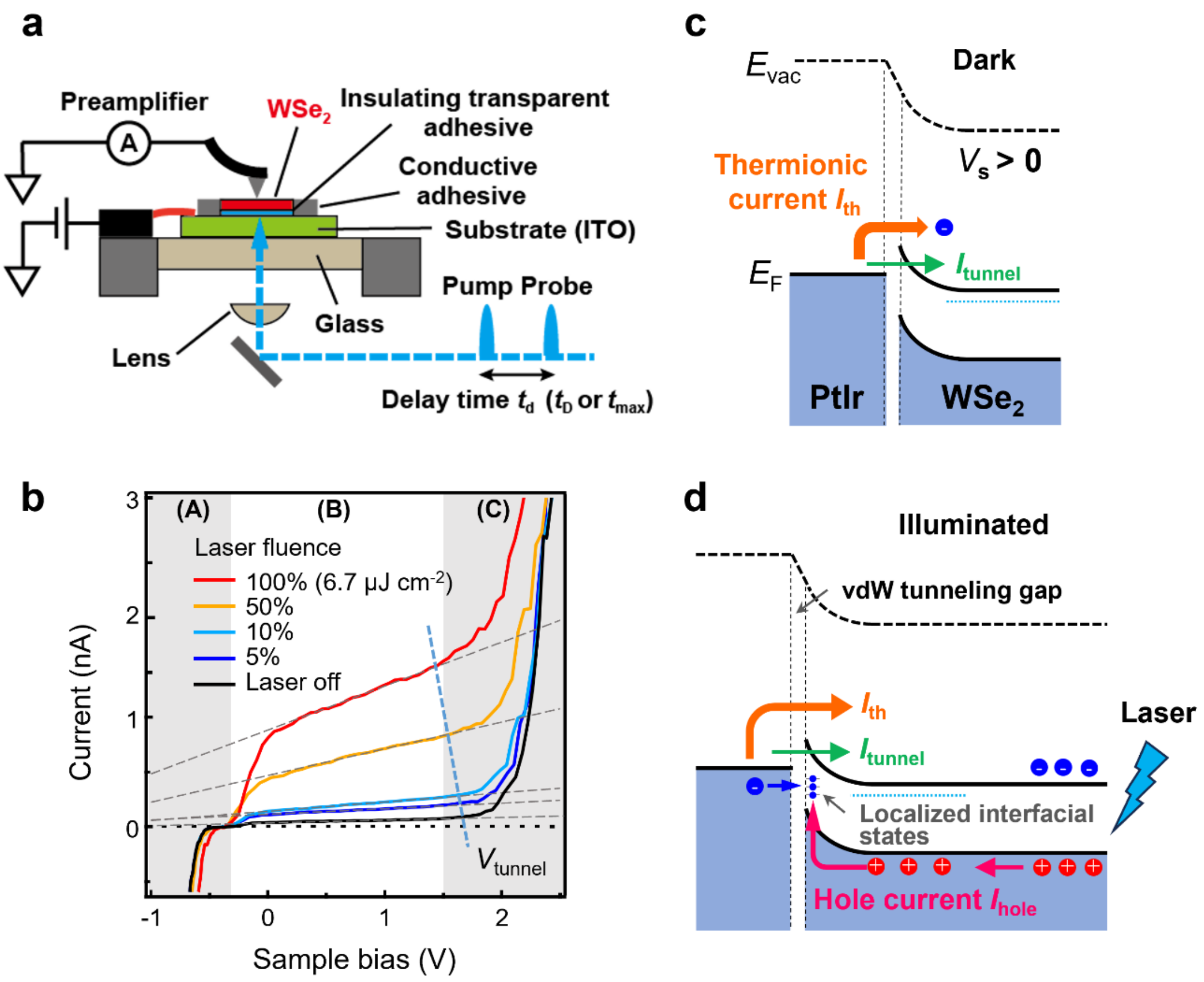
a
Preamplifier
A
WSe2
Insulating transparent adhesive
Conductive adhesive
Substrate (ITO)
Pump Probe
Glass
Lens
Delay time td (tD or tmax)
b
(A)
(B)
(C)
Laser fluence
100% (6.7 μJ cm-2)
50%
10%
5%
Laser off
Vtunnel
Current (nA)
Sample bias (V)
c
Evac
Dark
Vs > 0
Thermionic current Ith
EF
Itunnel
PtIr
WSe2
d
Illuminated
vdW tunneling gap
Ith
Itunnel
Laser
Localized interfacial states
Hole current Ihole

**Fig. 2**

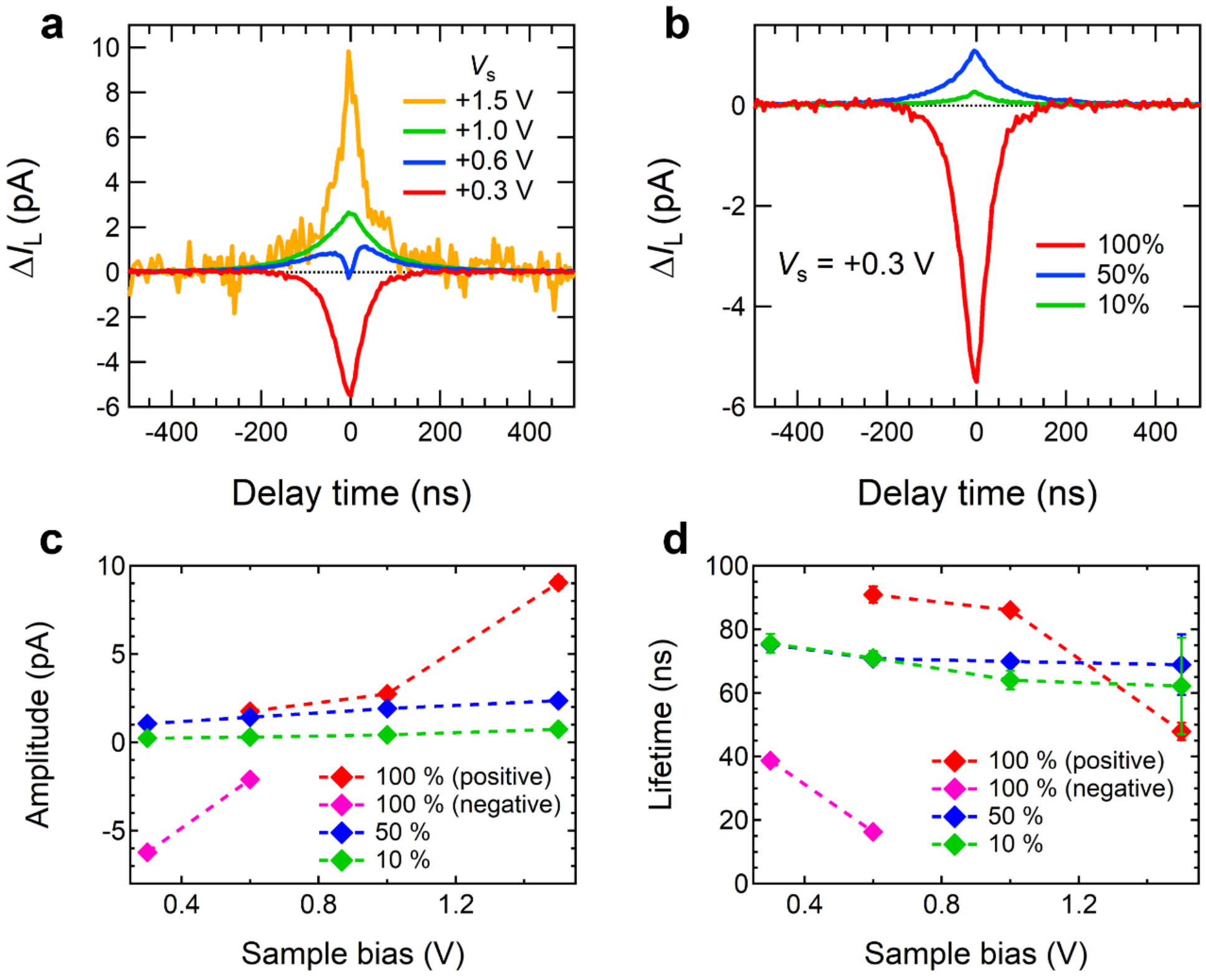

a
b
c
d
$V_s$
+1.5 V
+1.0 V
+0.6 V
+0.3 V
$\Delta I_L$ (pA)
Delay time (ns)
$V_s$ = +0.3 V
100%
50%
10%
Amplitude (pA)
Lifetime (ns)
Sample bias (V)
100 % (positive)
100 % (negative)
50 %
10 %

**Fig. 3**

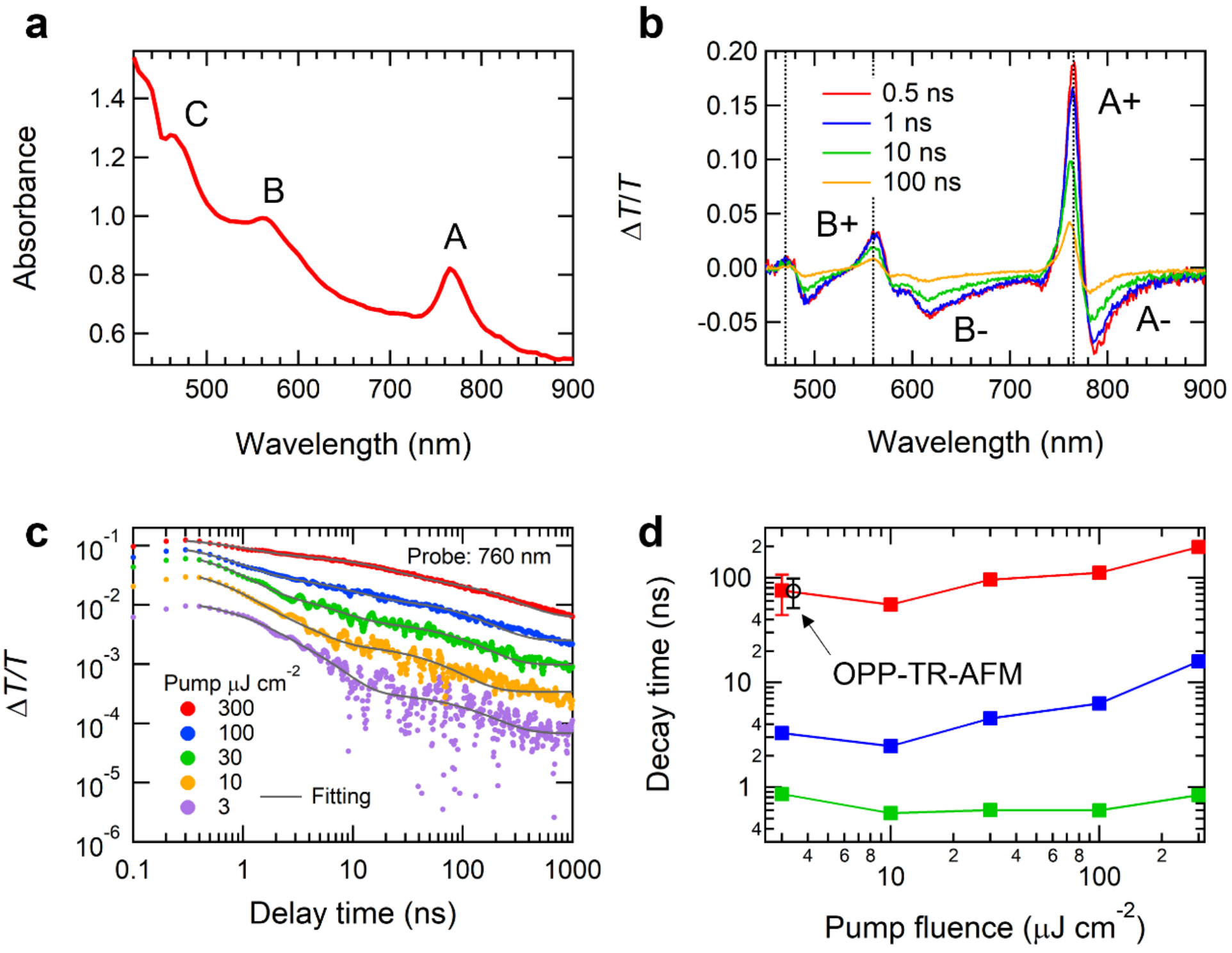
a
Absorbance
Wavelength (nm)
C
B
A
b
ΔT/T
0.5 ns
1 ns
10 ns
100 ns
A+
B+
B-
A-
Wavelength (nm)
c
ΔT/T
Probe: 760 nm
Pump μJ cm-2
300
100
30
10
3
Fitting
Delay time (ns)
d
Decay time (ns)
OPP-TR-AFM
Pump fluence (μJ cm-2)

Fig. 4

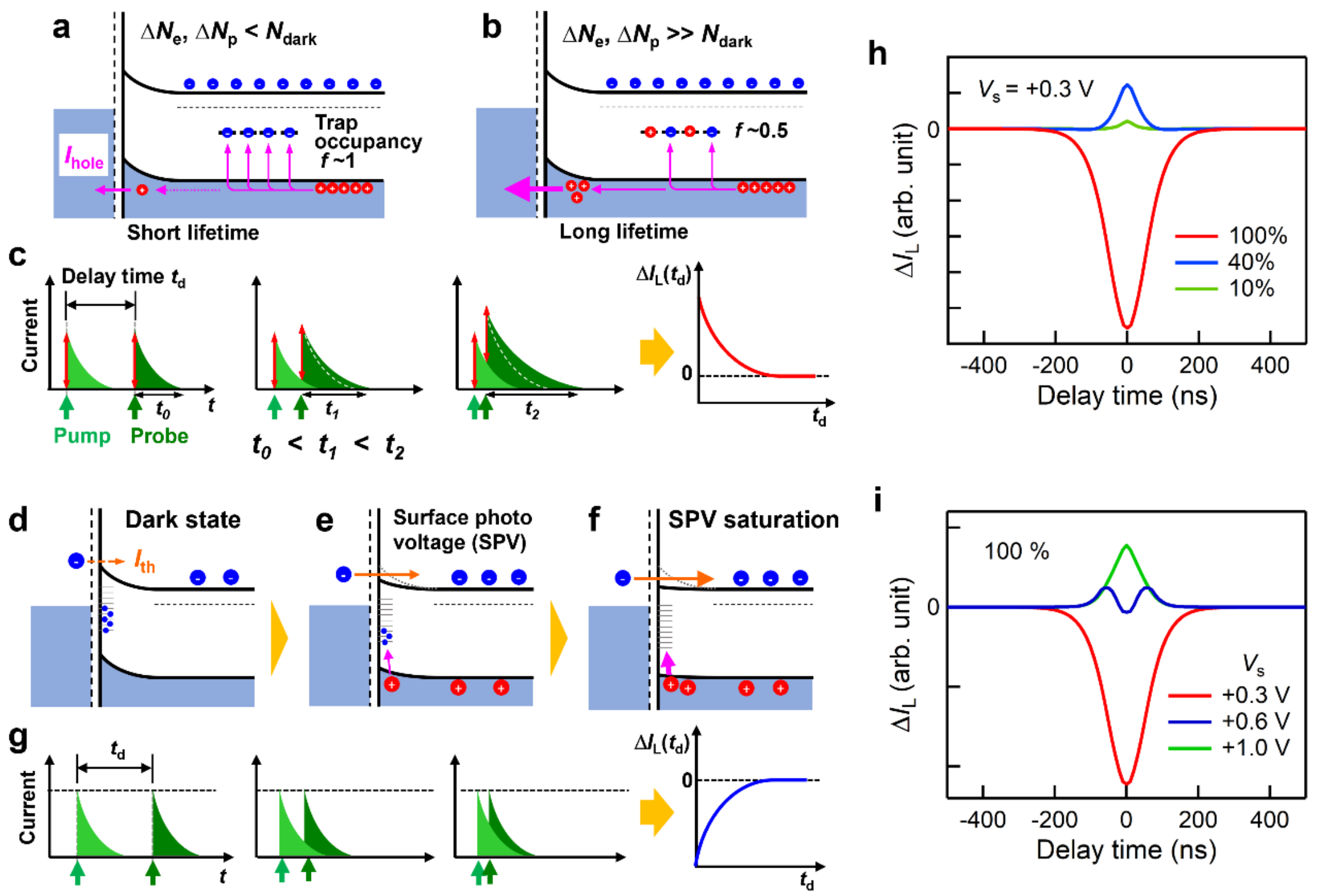

Fig. 5

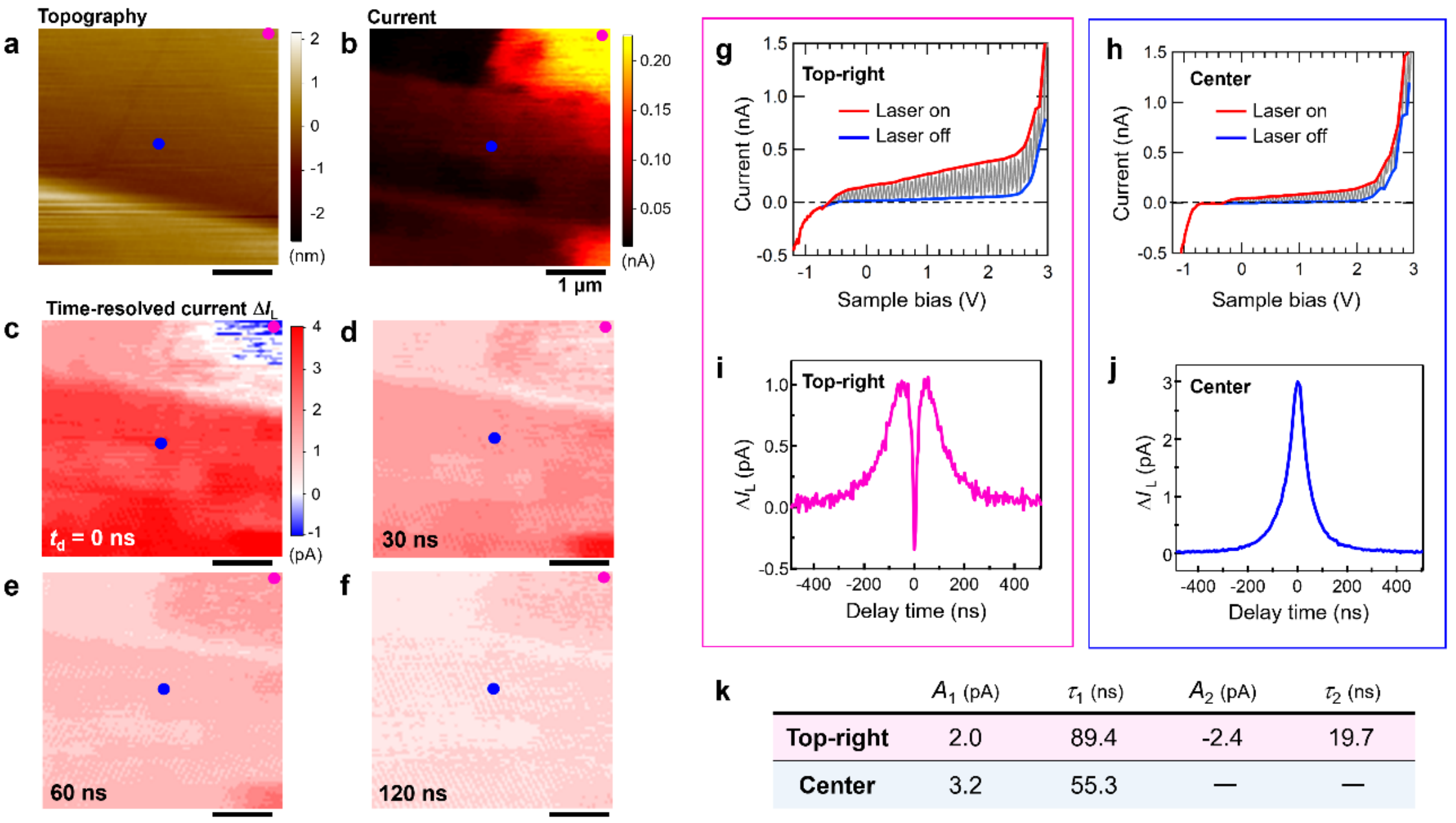

| | $A_1$ (pA) | $\tau_1$ (ns) | $A_2$ (pA) | $\tau_2$ (ns) |
|---|---|---|---|---|
| Top-right | 2.0 | 89.4 | -2.4 | 19.7 |
| Center | 3.2 | 55.3 | — | — |

**Supplementary Information**

# Intertwined bulk photocarrier and interfacial barrier dynamics in van der Waals point-contact Schottky junctions

Munenori Yokota[1,6], Hiroyuki Mogi[2,6], Yutaka Mera[3], Katsuya Iwaya[1*], Taketoshi Minato[4,5], Shoji Yoshida[2], Osamu Takeuchi[2], Tatsuo Nakagawa[1], and Hidemi Shigekawa[2*]

[1] UNISOKU Co., Ltd., Osaka 573-0131, Japan.

[2] Faculty of Pure and Applied Sciences, University of Tsukuba, Ibaraki 305-8573, Japan.

[3] Department of Fundamental Biosciences (Physics), Shiga University of Medical Science, Shiga 520-2192, Japan.

[4] Institute for Molecular Science (IMS), National Institutes of Natural Sciences, Aichi 444-8585, Japan.

[5] Core for Spin Life Sciences, Okazaki Collaborative Platform, National Institutes of Natural Sciences, Okazaki, Aichi 444-8585, Japan.

[6] These authors contributed equally: Munenori Yokota, Hiroyuki Mogi.

Corresponding author: Katsuya Iwaya (katsuya_iwaya@unisoku.co.jp)
Hidemi Shigekawa (hidemi@ims.tsukuba.ac.jp)

## 1. Optical pump–probe time-resolved atomic force microscopy (OPP-TR-AFM)

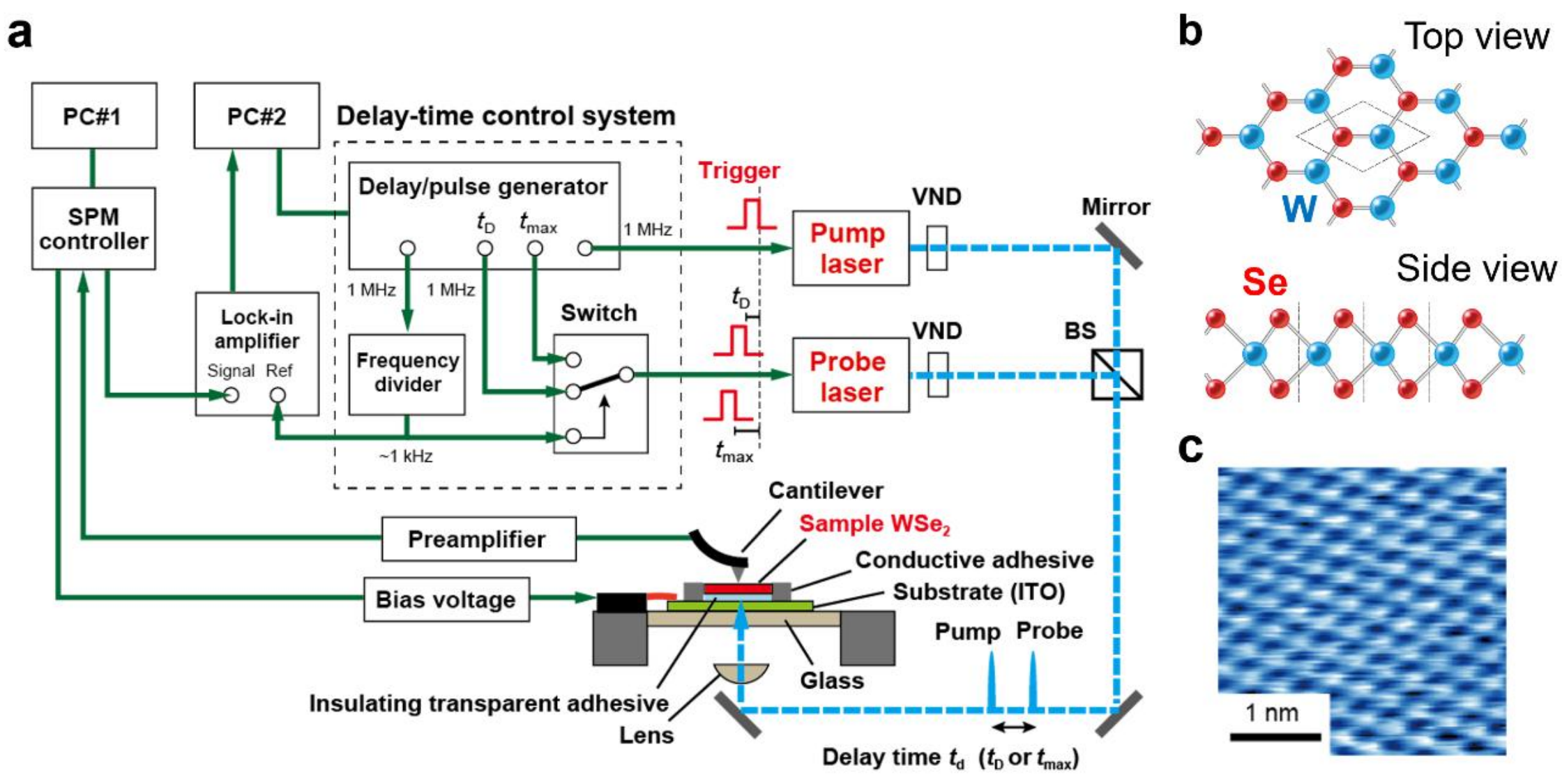

**Fig. S1: OPP-TR-AFM setup and $WSe_2$ sample characterization.**

(a) Schematic illustration of the OPP-TR-AFM system. BS: beam splitter, VND: variable neutral density filter, PC#1 and PC#2: control computers, $t_d$: delay-time variable, $t_D$: selected delay time, $t_{max}$: sufficiently long delay time used for the delay-time modulation method. (b) Crystal structure of $WSe_2$. (c) STM image of the $WSe_2$ surface.

## 2. Schottky barrier height estimation

The effective Schottky barrier height $\phi_B$ was estimated from the local dark *I*–*V* characteristics based on the thermionic emission model. Figure S2a shows the local dark *I*–*V* characteristics of the $WSe_2$/PtIr junction. The polarity of the rectifying behavior indicates that the $WSe_2$ surface behaves as an *n*-type semiconductor. Therefore, the negative-bias region, where the current exhibits an exponential increase, was analyzed using the thermionic emission model. A ln(-*I*)–*V* plot was constructed using the absolute value of the current, and the linear region was fitted to estimate the effective Schottky barrier height (Fig. S2b).

The saturation current $I_0$ was obtained from the intercept of the linear fit, and the Schottky barrier height was estimated according to the method of Giannazzo *et al.* [1] using

$$\phi_B = \frac{kT}{q}\ln\left(\frac{AA^*T^2}{I_0}\right) \qquad \text{(S1)}$$

where $k$ is the Boltzmann constant, $T$ is the absolute temperature, $q$ is the elementary charge, $A$ is the effective contact area, and $A^*$ is the effective Richardson constant.

Here, the contact area was estimated by assuming a tip radius of $r_{\mathrm{tip}}$ = 10 nm, $A = \pi r_{\mathrm{tip}}^2 = 3.14 \times 10^{-12}\ \mathrm{cm}^2$. The Richardson constant was taken to be $A^* = 35\text{–}50\ \mathrm{A\ cm^{-2}\ K^{-2}}$ based on the effective electron mass of $WSe_2$, and the calculation was performed at room temperature (300 K).

From the linear fitting of the $\ln(-I) - V$ plot, an intercept of $\ln I_0 = -36.771$ was obtained, yielding an estimated effective Schottky barrier height of $\phi_B = 0.65\text{–}0.66$ eV.

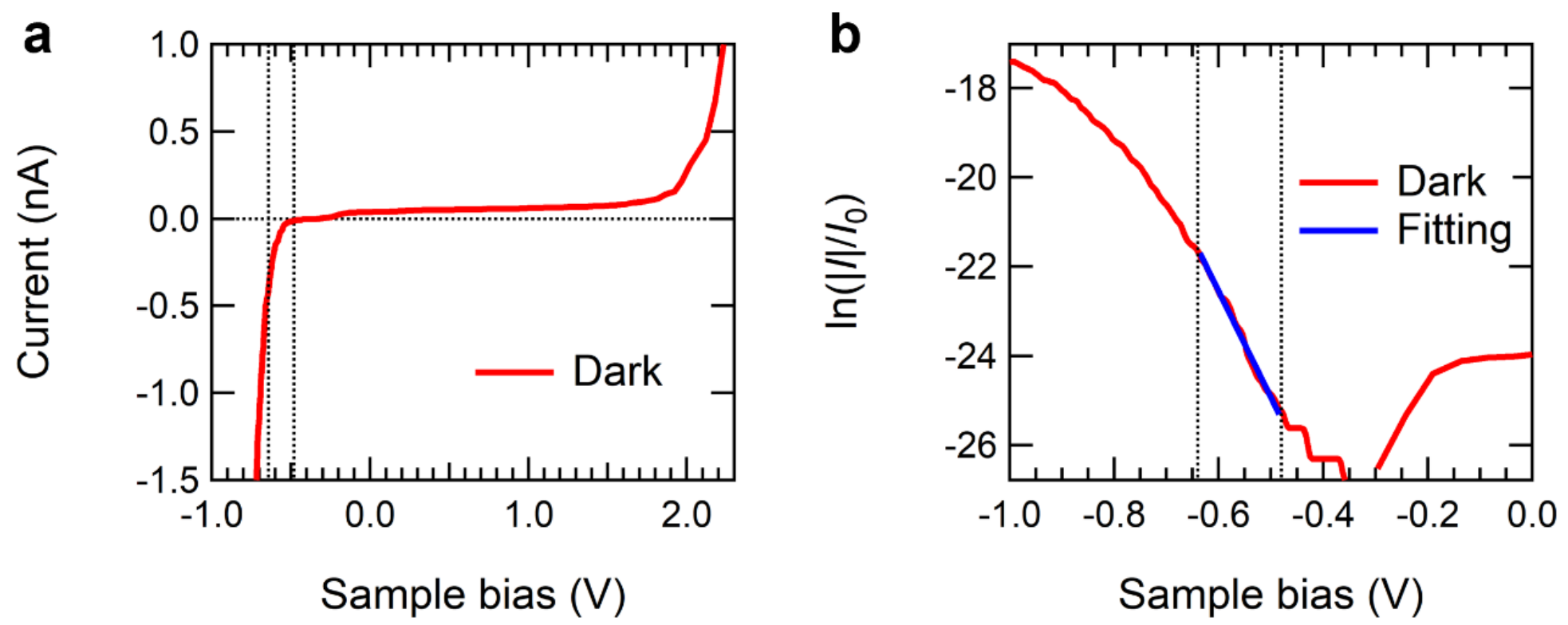

**Fig. S2: Estimation of the effective Schottky barrier height.**

(a) Local dark *I*–*V* characteristics of the $WSe_2$/PtIr junction. The effective Schottky barrier height was estimated using the thermionic emission model over the voltage range indicated by the dashed lines.

(b) ln(|*I*|/$I_0$)–*V* plot and the corresponding linear fit used to estimate the effective Schottky barrier height. $I_0$ = 1 A.

### 3. Contact force dependence of the time-resolved current

To evaluate the influence of mechanical contact on the OPP-TR-AFM signal and the structural stability of the junction, time-resolved current measurements were performed under different contact forces (Fig. S3). The contact force was varied among 12.0, 16.8, 21.6, and 26.4 nN. As the contact force increased, the amplitude of the time-resolved current increased, whereas no significant changes were observed in the temporal response profile or the relaxation behavior. These results indicate that increasing the contact force primarily enlarges the effective contact area, thereby increasing the measured current, while the carrier transport and interfacial barrier modulation dynamics in the van der Waals Schottky junction remain essentially unchanged over the investigated force range, demonstrating that a stable junction structure is maintained throughout the measurements.

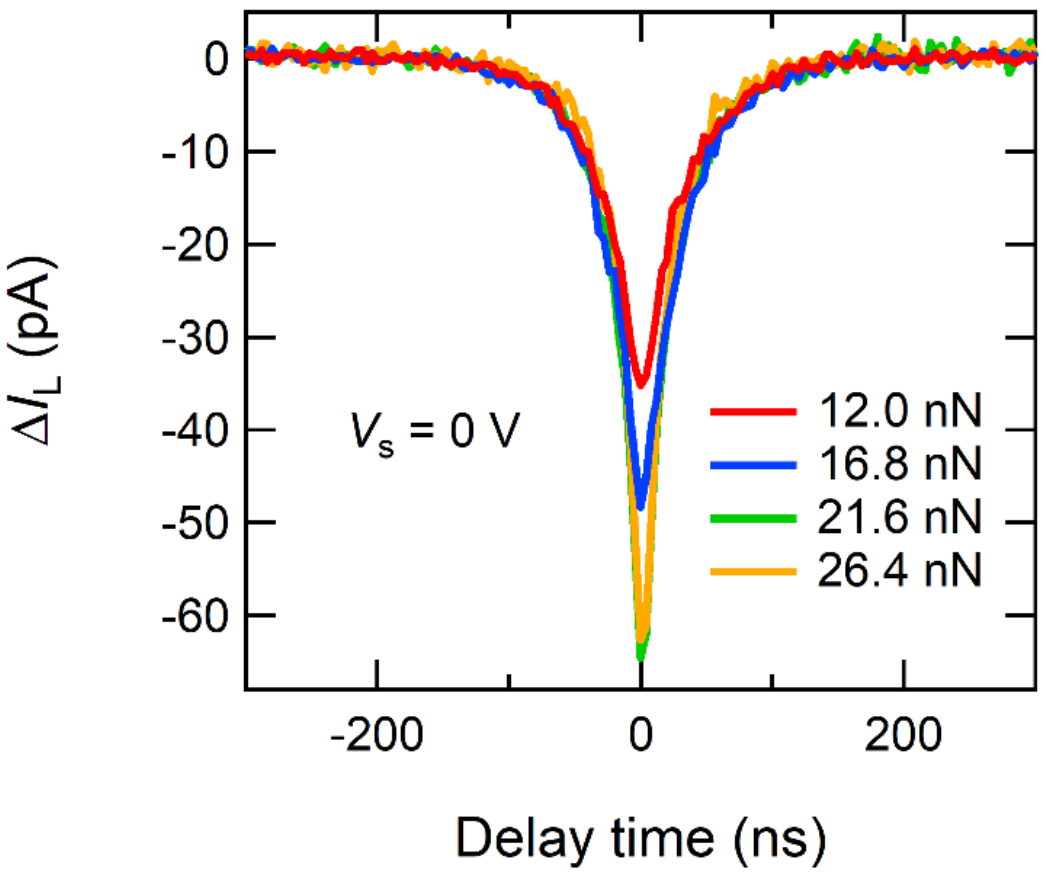

**Fig. S3: Contact force dependence of the time-resolved current.**

Time-resolved current measured under different contact forces at an excitation fluence of 3.0 μJ $cm^{-2}$ and $V_s$ = 0 V.

**4. Bias dependence of the average current and time-resolved current**

To further analyze the bias dependence of the time-resolved current observed in Fig. 2a of the main text, we compared the bias dependence of the average current under illumination, measured without delay-time modulation, with that of the time-resolved current $\Delta I_L(t_d)$ at a delay time of $t_d = 0$ (Fig. S4a).

In the forward-bias region (Region A), the average current under illumination exhibited a finite value, whereas $\Delta I_L(t_d = 0)$ remained nearly zero. This behavior is attributed to the fact that, under forward bias, hole accumulation near the interface and the associated surface photovoltage (SPV) are negligible, so that the photoexcited carriers primarily contribute to the steady-state photocurrent.

In contrast, in the reverse-bias region (Region B), $\Delta I_L(t_d = 0)$ reached its largest negative value at approximately +0.3 V. With increasing sample bias, the signal gradually increased and changed sign at approximately +0.6 V, becoming positive. In the tunneling-conduction region (Region C), both the illuminated average current and $\Delta I_L(t_d = 0)$ increased rapidly.

This bias dependence indicates that the contribution of SPV-induced barrier modulation becomes significant under positive sample bias. Furthermore, the sign reversal observed around +0.6 V demonstrates that the time-resolved current cannot be explained by a polarity reversal of a single current component; rather, it results from a change in the relative contributions of the negative and positive components as a function of sample bias. This interpretation is further supported by the time-resolved current traces shown in Fig. S4b. At low bias, the negative component dominates, whereas at +0.6 V both positive and negative components coexist. At higher bias, the positive component becomes dominant.

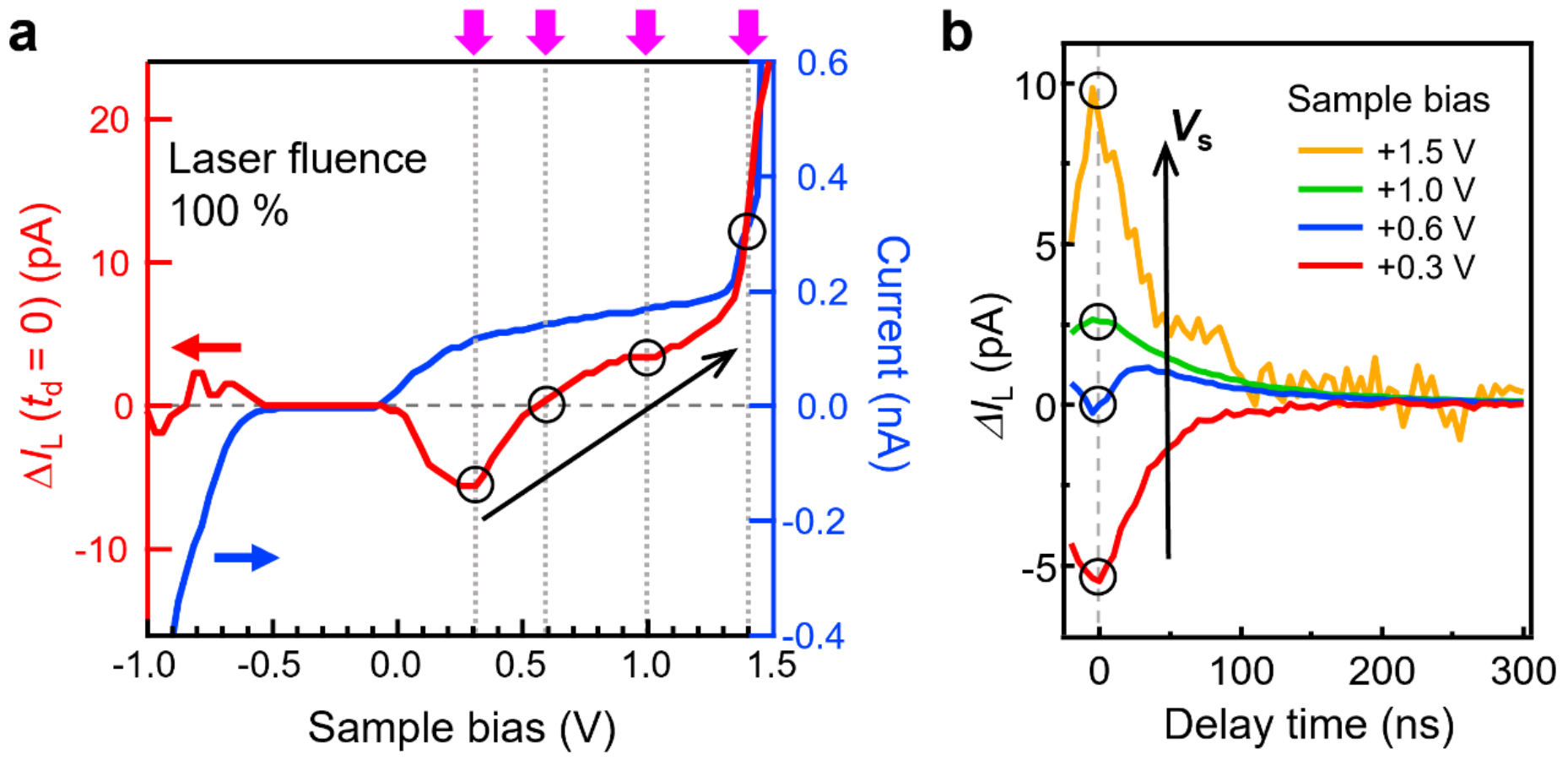

**Fig. S4: Bias dependence of the average current and time-resolved current.**

(a) Bias dependence of the average current under illumination (blue) and the time-resolved current $\Delta I_L(t_d = 0)$ (red). The black circles indicate the values of $\Delta I_L(t_d = 0)$ corresponding to the time-resolved traces shown in (b).

(b) Delay-time dependence of the time-resolved current measured at the sample biases indicated by the black circles in (a). At low bias, the negative component dominates the time-resolved response. With increasing sample bias, a positive component gradually emerges, and both components coexist at approximately +0.6 V. At higher bias, the positive component becomes dominant.

## 5. Excitation-power dependence of the photocurrent

To evaluate the nonlinearity of the photocurrent response, the excitation-power dependence was analyzed. Figure S5(a) shows the light-modulated *I*–*V* characteristics measured at different excitation fluences (the same data as those presented in Fig. 1b of the main text). The photocurrent ($I_{ph} = I_{ON} - I_{OFF}$) was extracted at each excitation fluence, and the resulting log–log plots are shown in Fig. S5(b). The photocurrent at $V_s = +0.3$, +1, and +2 V was fitted using a power-law relation, $I_{ph} \propto P^{\alpha}$ where $P$ denotes the excitation fluence, yielding $\alpha \approx 0.83$.

This analysis confirms that the photocurrent exhibits a sublinear dependence on excitation fluence under the present experimental conditions. The result indicates that the junction exhibits a nonlinear photoresponse and provides important background information for interpreting the time-resolved photocurrent response. However, the time-resolved current measured by OPP-TR-AFM reflects transient interfacial processes involving the transport of photocarriers generated in the bulk, charge accumulation at the interface, and modulation of the Schottky barrier. Therefore, the observed dynamics should be interpreted not only in terms of the excitation-power dependence but also in conjunction with their delay-time and bias dependences.

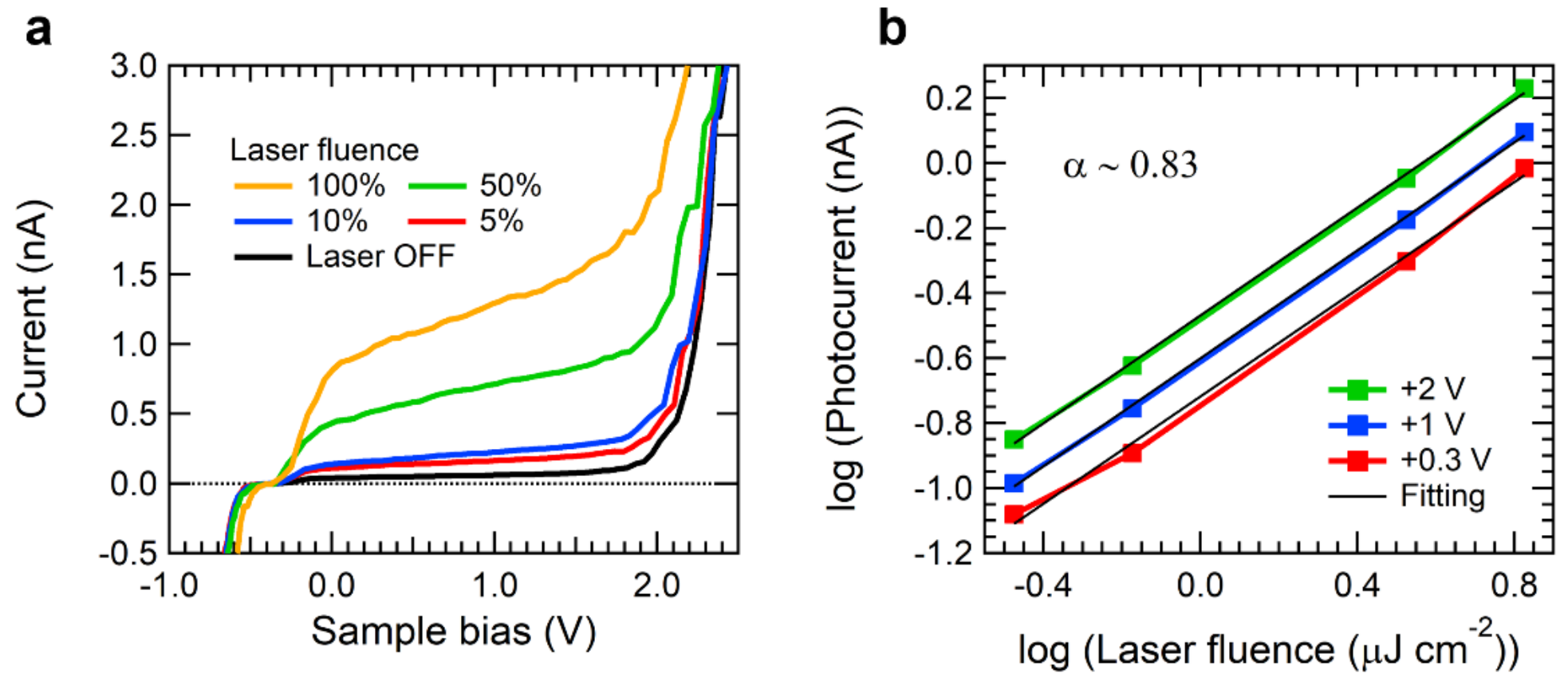

**Fig. S5: Power-law analysis of the excitation fluence dependence of the photocurrent.** (a) Light-modulated *I*–*V* characteristics measured at different excitation fluences (the same data as those presented in Fig. 1b of the main text). The excitation fluence corresponding to 100% is 6.7 μJ cm$^{-2}$.

(b) Log–log plots of the photocurrent as a function of excitation fluence at different sample bias voltages. The gray lines represent power-law fits using $I_{\text{ph}} \propto P^{\alpha}$, yielding $\alpha \approx 0.83$.

## 6. Transient absorption spectroscopy of bulk $WSe_2$

To investigate the photoexcited carrier dynamics in bulk $WSe_2$, transient absorption spectroscopy was performed using the RIPT method [2]. Fig. S6a illustrates the electronic band structure of bulk $WSe_2$ together with the optical transitions discussed in the main text. The A and B transitions originate from optical transitions from the spin–orbit-split valence bands (V1 and V2) at the K point to the conduction-band minimum (C1) [3]. In contrast, the higher-energy C transition corresponds to the band-nesting region. The 488 nm pump pulse (2.54 eV) used in this study predominantly excites carriers through the C-band transition.

Fig. S6b shows the delay-time dependence of the differential transmittance $\Delta T/T$ measured at a pump fluence comparable to that used in the OPP-TR-AFM experiments. Distinct spectral features associated with the excited-state dynamics near the A exciton resonance ($A^+$, $A^-$, and $A_2^-$) and the B exciton resonance ($B^+$) are clearly observed. The $A^+$ feature persists over a broad time window (0.1–1000 ns), as further confirmed by the $\Delta T/T$ spectra averaged over delay-time ranges of 0.4 ns, 10–100 ns, and 100–1000 ns, shown in Fig. S6c.

To quantitatively evaluate the carrier relaxation dynamics, the transient signals were analyzed using a phenomenological triple-exponential model. The pump-fluence dependences of the fitted amplitudes and lifetimes are presented in Figs. S6d and S6e, respectively. Three characteristic relaxation components were consistently identified over the entire pump-fluence range, providing the basis for comparison with the OPP-TR-AFM lifetime analysis discussed in the main text.

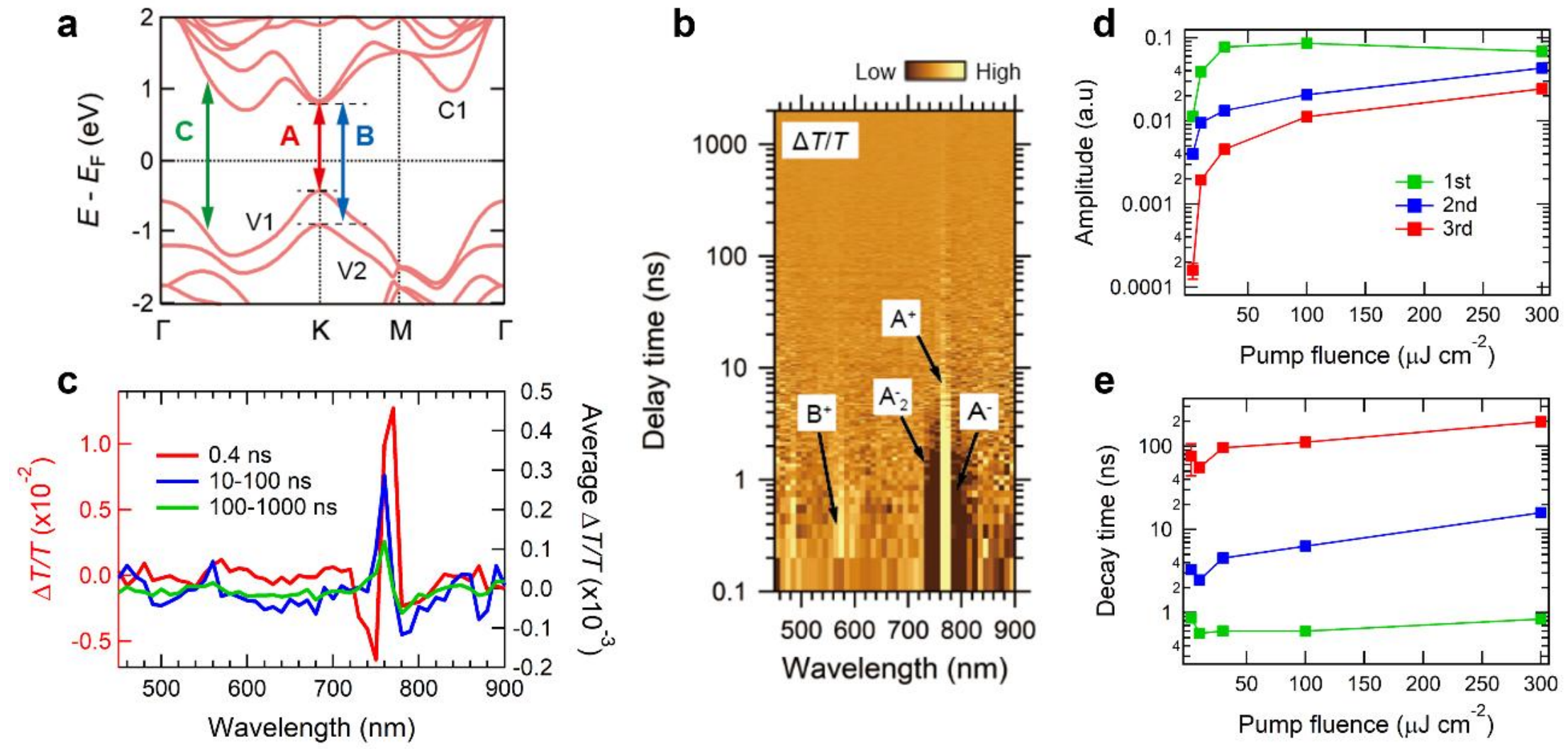

**Fig. S6: Transient absorption spectroscopy of bulk $WSe_2$.**

(a) Electronic band structure of bulk $WSe_2$ illustrating the optical transitions discussed in the main text. The A and B transitions originate from optical transitions from the spin–orbit-split valence bands (V1 and V2) at the K point to the conduction-band minimum (C1). The higher-energy C transition corresponds to the band-nesting region.

(b) Differential transmittance ($\Delta T/T$) map measured following 488 nm excitation at a pump fluence of 3 μJ cm⁻², comparable to that used in the OPP-TR-AFM measurements. The spectral features are assigned to $A^+$, $A^-$, $A_2^-$, and $B^+$.

(c) $\Delta T/T$ spectra extracted from (b) at a delay time of 0.4 ns and averaged over the delay-time ranges of 10–100 ns and 100–1000 ns.

(d, e) Pump-fluence dependences of the amplitudes (d) and lifetimes (e) of the $A^+$ component at 760 nm, obtained by triple-exponential fitting.

## 7. Analysis of thermal and non-thermal contributions to the transient absorption spectra

To evaluate the contribution of laser heating to the long-lived response observed in the TAS measurements, a fitting model was employed to quantitatively analyze the transient absorbance spectra ($\Delta A$) at a delay time of 10 ns.

First, the region around the A excitonic resonance in the steady-state absorbance spectrum was fitted using the sum of a Lorentzian function representing the excitonic resonance and a second-order polynomial background,

$$A(E) = aE^2 + bE + c + \frac{I_0 \Gamma_0^2}{(E-E_0)^2+\Gamma_0^2} \quad \text{(S2)}$$

where $E$ is the photon energy; $a$, $b$, and $c$ are the coefficients describing the quadratic background; $E_0$ is the excitonic resonance energy; $\Gamma_0$ is the half width at half maximum (HWHM) of the excitonic resonance; and $I_0$ is the peak intensity.

Next, using the fitted steady-state absorption spectrum as the reference, we defined a modified absorption spectrum incorporating an excitonic energy shift $E_{\mathrm{shift}}$, linewidth broadening $E_{\mathrm{broad}}$, and amplitude change $I_{\mathrm{coef}}$:

$$A'(E) = a(E - E_{\mathrm{shift}})^2 + b(E - E_{\mathrm{shift}}) + c + \frac{I_{\mathrm{coef}} I_0 (\Gamma_0 + E_{\mathrm{broad}})^2}{(E-(E_0+E_{\mathrm{shift}}))^2+(\Gamma_0+E_{\mathrm{broad}})^2} \quad \text{(S3)}$$

The transient absorbance spectrum was then calculated as

$$\Delta A(E) = A'(E) - A(E) \quad \text{(S4)}$$

and the experimentally measured $\Delta A$ spectrum at a delay time of 10 ns was fitted using $E_{\mathrm{shift}}$, $E_{\mathrm{broad}}$, and $I_{\mathrm{coef}}$ as the fitting parameters.

Figure S7a compares the measured $\Delta A$ spectra around the A excitonic resonance (dashed lines) with the corresponding fitted spectra (solid lines) for each pump fluence. Despite the small number of fitting parameters, the calculated differential spectra reproduce the experimental spectra well.

Figures S7b and S7c show the pump-fluence dependence of the extracted excitonic energy shift $E_{\mathrm{shift}}$ and additional linewidth broadening $E_{\mathrm{broad}}$, respectively. Both the energy shift and linewidth broadening increase with increasing pump fluence, although the linewidth broadening becomes more pronounced under high-excitation conditions.

Arora *et al.* reported that, for bulk $WSe_2$ near room temperature, both the excitonic energy shift and linewidth broadening in the steady-state absorbance spectrum increase approximately linearly with increasing temperature [4]. Based on their reported temperature dependence, the expected ratio for a purely thermal response is

$$|E_{\mathrm{shift}}|/E_{\mathrm{broad}} \approx 2.3 \quad \text{(S5)}$$

Figure S7d shows the extracted $|E_{\mathrm{shift}}|/E_{\mathrm{broad}}$ ratio at each pump fluence. Under the low-excitation conditions corresponding most closely to those employed in the OPP-TR-AFM measurements, the ratio is approximately 1, indicating that the linewidth broadening is relatively larger than the energy shift. Since exciton–carrier scattering is known to broaden excitonic resonances without inducing a significant energy shift [5], the excess linewidth broadening suggests that interactions with photoexcited carriers persisting beyond 10 ns play a dominant role under these conditions.

At higher pump fluences, the ratio obtained at 100 μJ $cm^{-2}$ approaches the thermal expectation, indicating a larger contribution from photoinduced heating. However, the ratio decreases again at the highest pump fluence of 300 μJ $cm^{-2}$ rather than remaining close to the thermal value. Moreover, as shown in Fig. 3d of the main text, the carrier lifetime increases with increasing pump fluence, which is expected to increase the density of photoexcited carriers remaining at a delay time of 10 ns. These results suggest that, although photoinduced heating contributes to the transient spectral evolution, it cannot fully account for the observed pump-fluence dependence, and interactions involving residual photoexcited carriers remain important even at the highest excitation level.

Taken together, these results indicate that long-lived photoexcited carriers remain present throughout the excitation range investigated, although their relative contribution varies with pump fluence. Under the low-excitation conditions employed for the OPP-TR-AFM measurements, carrier-related effects dominate over photoinduced heating. This supports the simulation model in which the long-lived time-resolved response originates from the diffusion of photocarriers generated within bulk $WSe_2$.

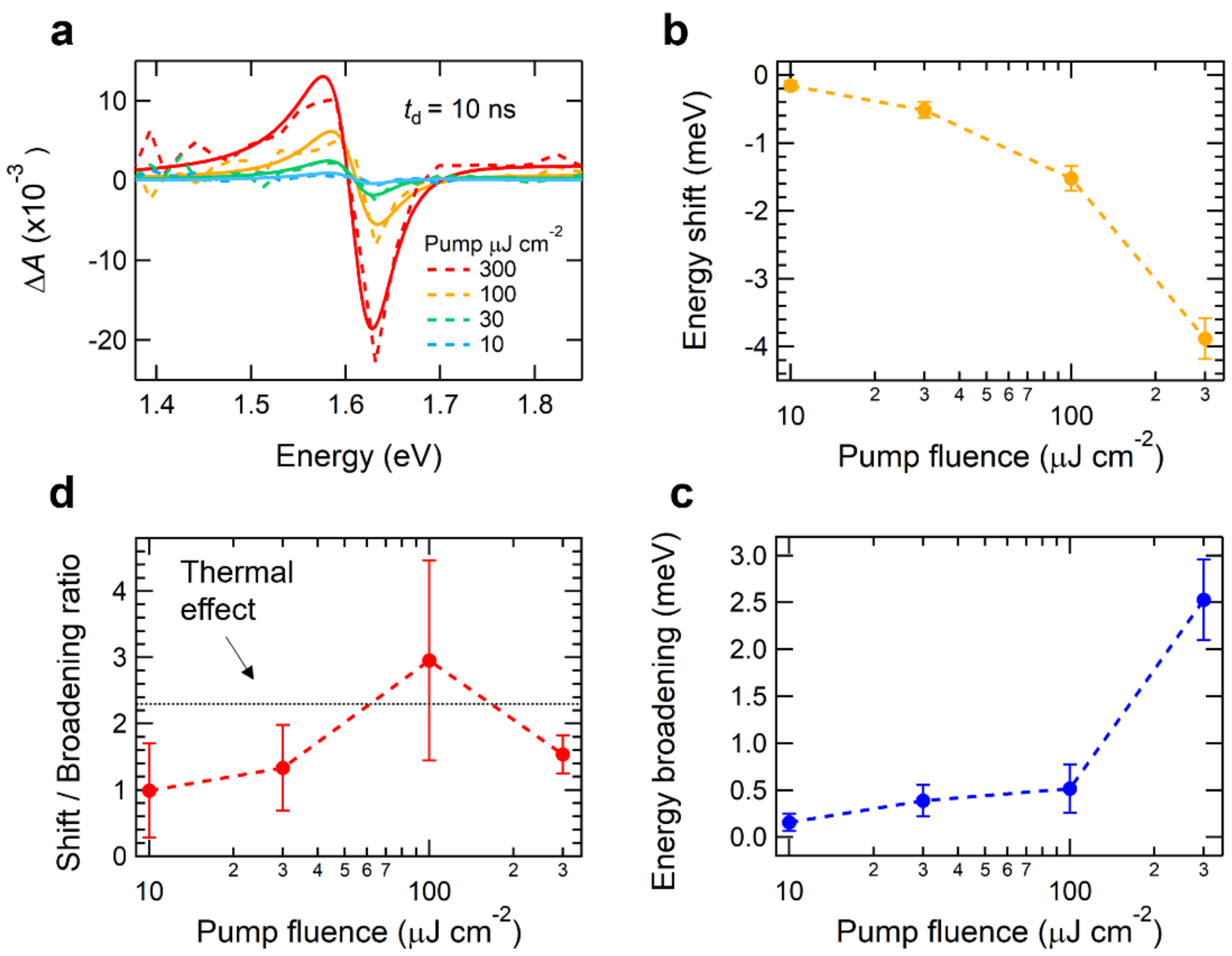

**Fig. S7: Evaluation of the thermal contribution to the transient absorbance spectra.** (a) Measured transient absorbance spectra, $\Delta A$ at $t_d$ = 10 ns (dashed lines), and the corresponding fitting results (solid lines) for different pump fluences. The spectra were fitted using a differential spectrum model based on the steady-state absorption spectrum, in which the excitonic energy shift, linewidth broadening, and amplitude change were treated as fitting parameters. (b) Pump-fluence dependence of the extracted excitonic energy shift $E_{shift}$. (c) Pump-fluence dependence of the extracted additional linewidth broadening $E_{broad}$. (d) Pump-fluence dependence of $|E_{shift}|/E_{broad}$. The black dashed line indicates the expected value ~2.3 for a purely thermal response estimated from the temperature dependence of bulk $WSe_2$ reported by Arora et al. [4]. Under low-excitation conditions, the ratio falls below the thermal expectation, suggesting that the observed spectral evolution cannot be explained solely by a temperature increase.

## 8. Calculation of the hole lifetime based on the SRH model

To support the hole diffusion current model proposed in the main text, the dependence of the effective hole lifetime on the photoexcited carrier density was calculated using the Shockley–Read–Hall (SRH) recombination model, taking the trap-state occupancy into account [6]. Following the procedure described in Ref. 6, the dark electron density $N_{\mathrm{dark}}$ was treated as a parameter, while the photoexcited carrier density $\Delta N_{\mathrm{p}}(=\Delta N_{\mathrm{e}})$ was varied.

As shown in Fig. S8, the effective hole lifetime increases with increasing photoexcited carrier density and gradually saturates at high excitation densities. This behavior arises because increasing the photoexcited carrier density reduces the electron occupancy of the trap states, thereby suppressing trap-assisted recombination. These results are consistent with the interpretation presented in the main text that a higher excitation density allows more photogenerated holes to reach the interface, resulting in an increased hole diffusion current.

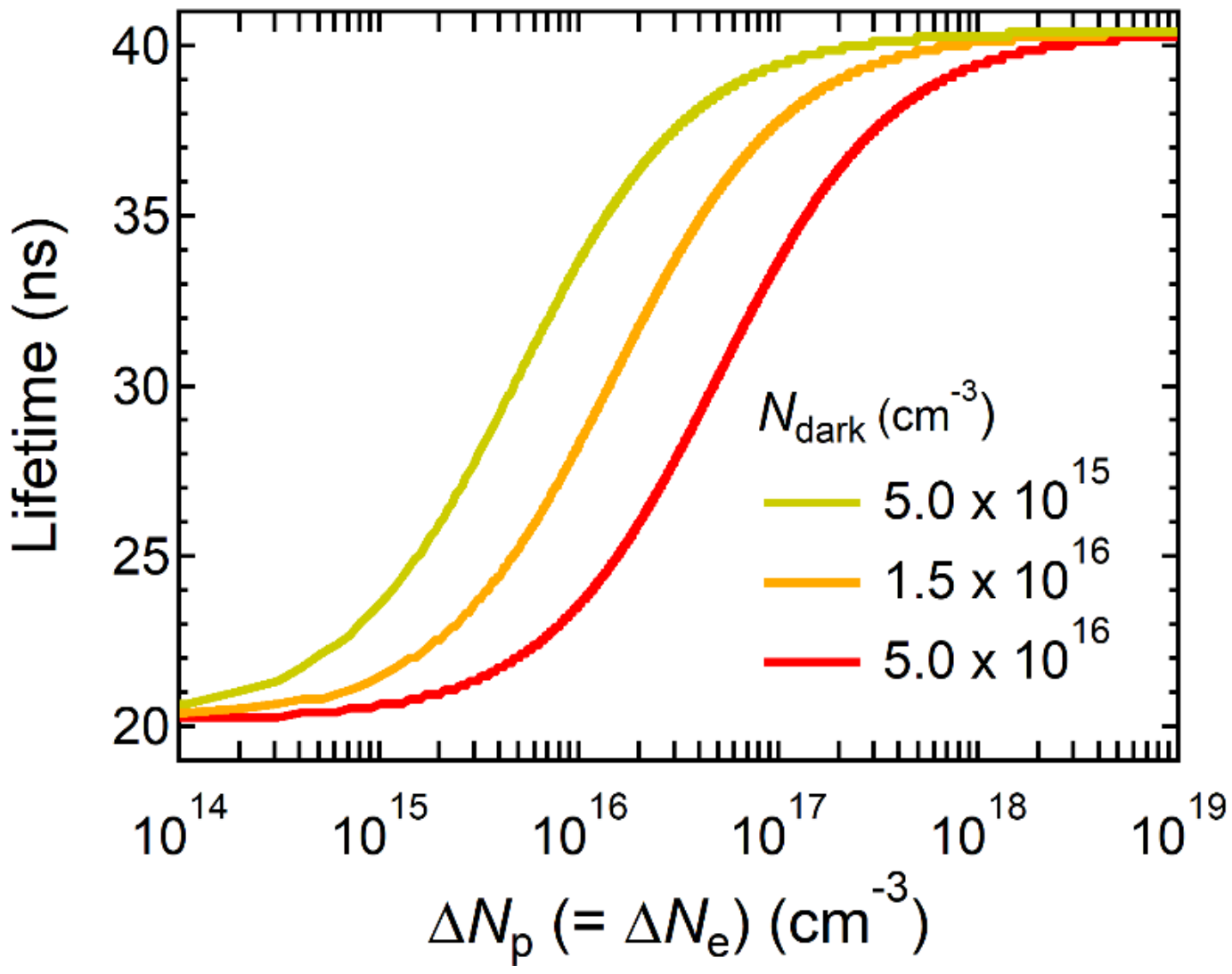

**Fig. S8: Effective hole lifetime calculated using the SRH recombination model.** Effective hole lifetime as a function of the photoexcited carrier density ($\Delta N_{\mathrm{p}} = \Delta N_{\mathrm{e}}$), calculated using the SRH recombination model with trap-state occupancy taken into account.

## 9. Numerical simulations of the time-resolved current response

To understand the positive and negative time-resolved current responses observed in the OPP-TR-AFM measurements, numerical simulations were performed based on the physical model summarized in Fig. 4 of the main text. The model considers two competing current contributions: (i) Hole diffusion current ($I_{\mathrm{hole}}$) arising from photogenerated holes in the bulk $WSe_2$ that diffuse toward the $WSe_2$/PtIr interface and (ii) Thermionic current ($I_{\mathrm{th}}$) modulated by photoinduced changes in the effective

Schottky barrier.

In the backside-illumination geometry employed in this study, photoexcited carriers are generated predominantly within the bulk $WSe_2$ region. Direct photoexcitation within the depletion region or the metal electrode is therefore assumed to make a negligible contribution. Accordingly, the experimentally observed time-resolved current is modeled as the superposition of the photoinduced changes in $I_{hole}$ and $I_{th}$.

**(1) Thermionic current**

The thermionic current $I_{th}$ is described by [7]

$$I_{th} = \exp\left(\frac{qV_{is}(t)}{kT}\right) I_{dark} \qquad \text{(S6)}$$

where $q$ is the elementary charge, $V_{is}(t)$ is the interface potential modulation induced by charge accumulated in localized interfacial states at the $WSe_2$/PtIr junction, $k$ is the Boltzmann constant, and $T$ is the absolute temperature.

The dark current is given by

$$I_{dark} = I_0 \left(\exp\left(-\frac{qV_R}{nkT}\right) - 1\right) \qquad \text{(S7)}$$

where $V_R$ is the applied reverse bias, $n$ is the ideality factor of the Schottky junction, and $I_0$ is the reverse saturation current. Here, the reverse-bias voltage $V_R$ corresponds to the positive sample bias $V_s$ applied during the OPP-TR-AFM measurements.

The interface potential was assumed to vary logarithmically with the population of holes trapped in the localized interfacial states [8]:

$$V_{is}(t) = V_{is0} \ln\left(1 + \alpha \frac{N_{is}(t)}{N_{dark}}\right) \qquad \text{(S8)}$$

where $N_{is}$ is the population of holes occupying the localized interfacial states, $N_{dark}$ is the equilibrium electron density under dark conditions, $V_{is0}$ determines the magnitude of the interface potential modulation, and $\alpha$ describes the efficiency with which hole accumulation in the localized interfacial states modifies the effective Schottky barrier height.

The holes occupying the localized interfacial states were assumed to relax through electron tunneling from the metal electrode:

$$N_{is}(t) = N_{is0} \cdot \exp\left(-\frac{t}{\tau_{is}}\right) \qquad \text{(S9)}$$

where $\tau_{is}$ is the characteristic relaxation time and $N_{is0}$ is a fitting parameter that determines the initial population of the localized interfacial states.

To reproduce the experimentally observed bias dependence of the time-resolved response, the relaxation time was assumed to decrease linearly with reverse bias according to

$$\tau_{\mathrm{is}} = \tau_{\mathrm{is0}}(1 - \beta V_R) \qquad \text{(S10)}$$

where $\tau_{\mathrm{is0}}$ is the relaxation time at zero bias and $\beta$ is a fitting parameter describing the reduction in relaxation time with increasing reverse bias.

In the simulations, $\tau_{\mathrm{is}}$ was set to 44 ns at $V_s = +0.3$ V and 27 ns at $V_s = +0.6$ V, which are close to the values obtained experimentally from the OPP-TR-AFM measurements (Fig. 2d in the main text).

**(2) Hole diffusion current**

As illustrated in Figs. 4a and 4b of the main text, the effective hole lifetime was assumed to depend on the occupancy of bulk trap states that govern SRH recombination.

Under low-excitation conditions ($\Delta N_{\mathrm{p}} \ll N_{\mathrm{dark}}$), a large number of empty bulk trap states are available, leading to efficient trapping of photogenerated holes and a relatively short effective lifetime. As the photoexcited carrier density increases, the trap states become progressively occupied, reducing the trapping probability and thereby increasing the effective hole lifetime toward a saturation value.

This carrier-density dependence is naturally described by the SRH recombination model when trap-state occupancy is taken into account. The calculated dependence of the effective hole lifetime on the photoexcited carrier density is shown in Fig. S8. To reproduce this behavior using a simple expression suitable for numerical simulations, we adopted the following phenomenological approximation:

$$\tau_{\mathrm{p}} = \tau_{\mathrm{p,base}} + \tau_{\mathrm{p,base}} \frac{1}{1 + N_{\mathrm{dark}}/N_{\mathrm{p}}} \qquad \text{(S11)}$$

where $\tau_{\mathrm{p,base}}$ is the hole lifetime in the low-excitation limit, $N_{\mathrm{p}}$ is the instantaneous photoexcited hole density, and $N_{\mathrm{dark}}$ is the equilibrium electron density under dark conditions.

In the simulations, the high-excitation lifetime was set to 40 ns, close to the value obtained from the TAS measurements (Fig. S6e), whereas the low-excitation lifetime was set to 20 ns, corresponding to half the high-excitation value.

The corresponding hole diffusion current was assumed to be proportional to the instantaneous photoexcited hole density,

$$I_{\mathrm{hole}} = qN_{\mathrm{p}}(t)\Lambda \qquad \text{(S12)}$$

where $N_p(t)$ is the time-dependent hole density and $\Lambda$ is a fitting parameter that incorporates both hole diffusion in the bulk and carrier capture into the localized interfacial states.

A longer hole lifetime leads to a slower decay of the bulk hole population. Consequently, the hole diffusion current $I_{\mathrm{hole}}$ (Eq. S12) persists for a longer period, continuously supplying holes to the localized interfacial states. This, in turn, influences both the occupancy of the localized interfacial states $N_{\mathrm{is}}$, and the interface potential modulation $V_{\mathrm{is}}$, thereby affecting the thermionic current $I_{\mathrm{th}}$.

**(3) Simulation procedure and decomposition of the time-resolved current**

Under these assumptions, each current component was calculated as a function of the delay time $t_{\mathrm{d}} = t_{\mathrm{D}}$. The total current under optical excitation $I_{\mathrm{L}}(t_{\mathrm{D}})$, was then obtained by summing these two current components. Finally, following the experimental procedure described by Eq. 1 in the main text, the time-resolved current response $\Delta I_{\mathrm{L}}(t_{\mathrm{D}})$ was calculated. For the high-excitation condition, the simulations were performed assuming that nearly all localized interfacial states were occupied by holes immediately after photoexcitation.

Although the hole diffusion current and the thermionic current were treated as separate components for simplicity, they are physically coupled. The interfacial charge accumulation responsible for the surface photovoltage modulation originates from photogenerated holes that diffuse from the bulk $WSe_2$ to the $WSe_2$/PtIr interface and become trapped in the localized interfacial states. Consequently, modulation of the thermionic current is ultimately driven by the same photogenerated holes that give rise to the hole diffusion current.

The simulation results demonstrate that the experimentally observed time-resolved current response arises from the superposition of the hole diffusion current and the thermionic current modulation. Fig. S9 presents the simulated time-resolved signal calculated for $V_s = +0.6$V under 100% excitation fluence, together with its decomposition into the hole diffusion and thermionic current components.

As indicated by the arrow in Fig. S9, the thermionic current component also exhibits a small positive contribution around a delay time of approximately 70 ns. This behavior arises because the thermionic current is influenced not only by the occupancy of the localized interfacial states but also by the time-dependent supply of photogenerated holes from the bulk to the interface. As holes continue to arrive at the interface, the formation and relaxation of the SPV deviate slightly from a simple monotonic recovery, giving rise to a weak positive contribution to the thermionic current component.

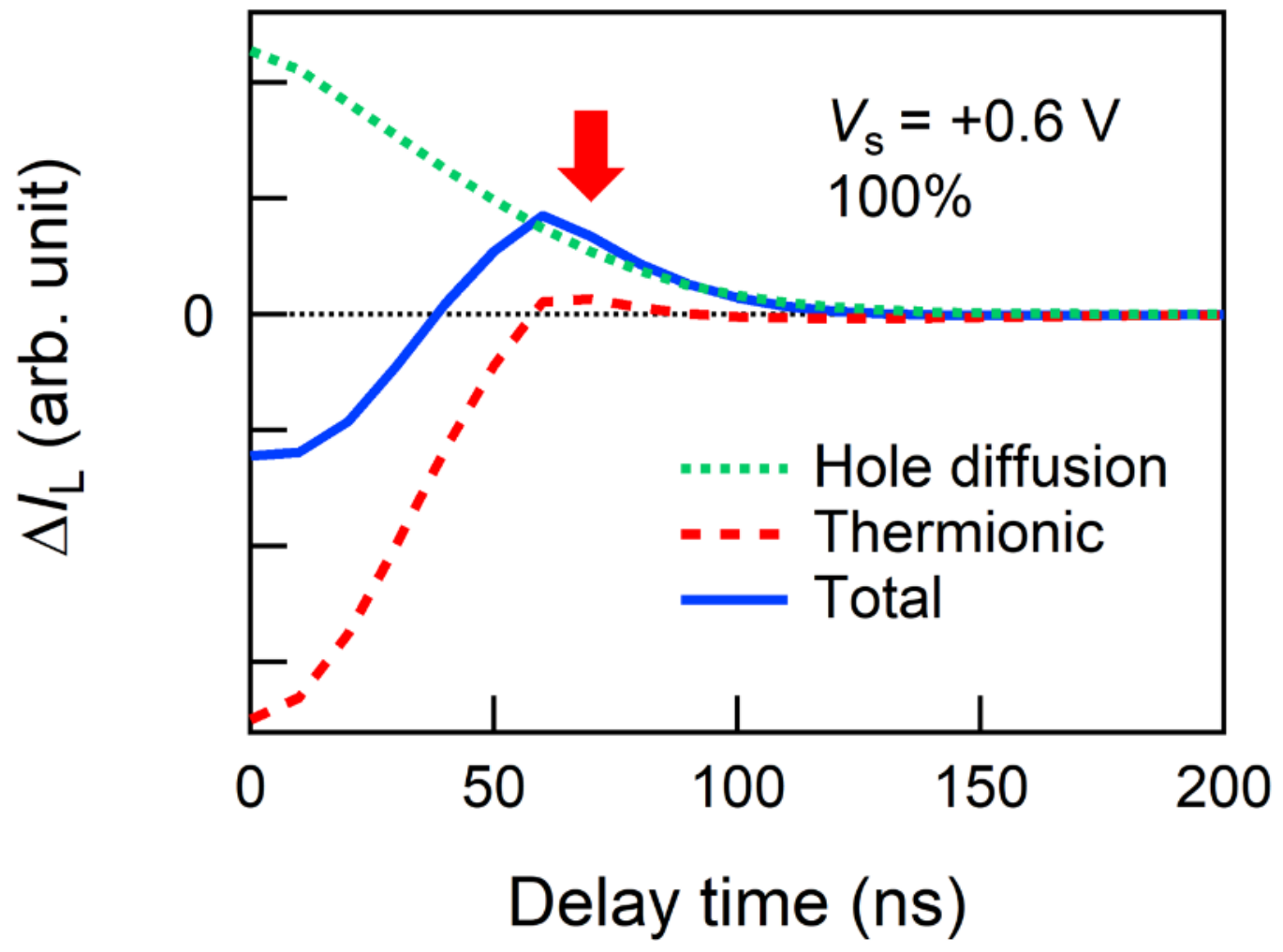

**Fig. S9: Decomposition of the simulated time-resolved current response.**
Decomposition of the simulated time-resolved current response calculated for $V_s = +0.6$ V and 100% excitation fluence (corresponding to the 100% excitation condition at $V_s = +0.6$ V in Fig. 2a of the main text) into the hole diffusion current and thermionic current components. As indicated by the red arrow, the thermionic current component also exhibits a weak positive contribution.

**10. Simulated temporal evolution of the photoinduced interfacial potential**

To visualize the temporal evolution of the effective interfacial potential introduced in the simulation model, the calculated interface potential modulation $V_{\mathrm{is}}(t)$ (Eq. S8) is shown in Fig. S10. As described in Section 9, the interface potential originates from holes accumulated in the localized interfacial states and modulates the effective Schottky barrier, thereby changing the thermionic current (Fig. S10a).

Figures S10b–d show the simulated evolution of $V_{\mathrm{is}}(t)$ for representative pump–probe delay times of $t_d = 50$, 100, and 350 ns under different sample-bias conditions. Because the relaxation time of the interface potential decreases with increasing reverse bias in the model, the interface potential recovers more rapidly at higher bias voltages. Consequently, under low-bias conditions, the interface potential remains only partially relaxed at short delay times, whereas it nearly recovers before the arrival of the probe pulse at longer delay times. In contrast, at higher bias voltages, the faster relaxation leads to a more rapid recovery of the interface potential.

These simulated results provide a direct visualization of the time-dependent modulation of the effective interfacial potential incorporated in the model and help explain the bias dependence of the thermionic-current component observed in the OPP-TR-AFM measurements.

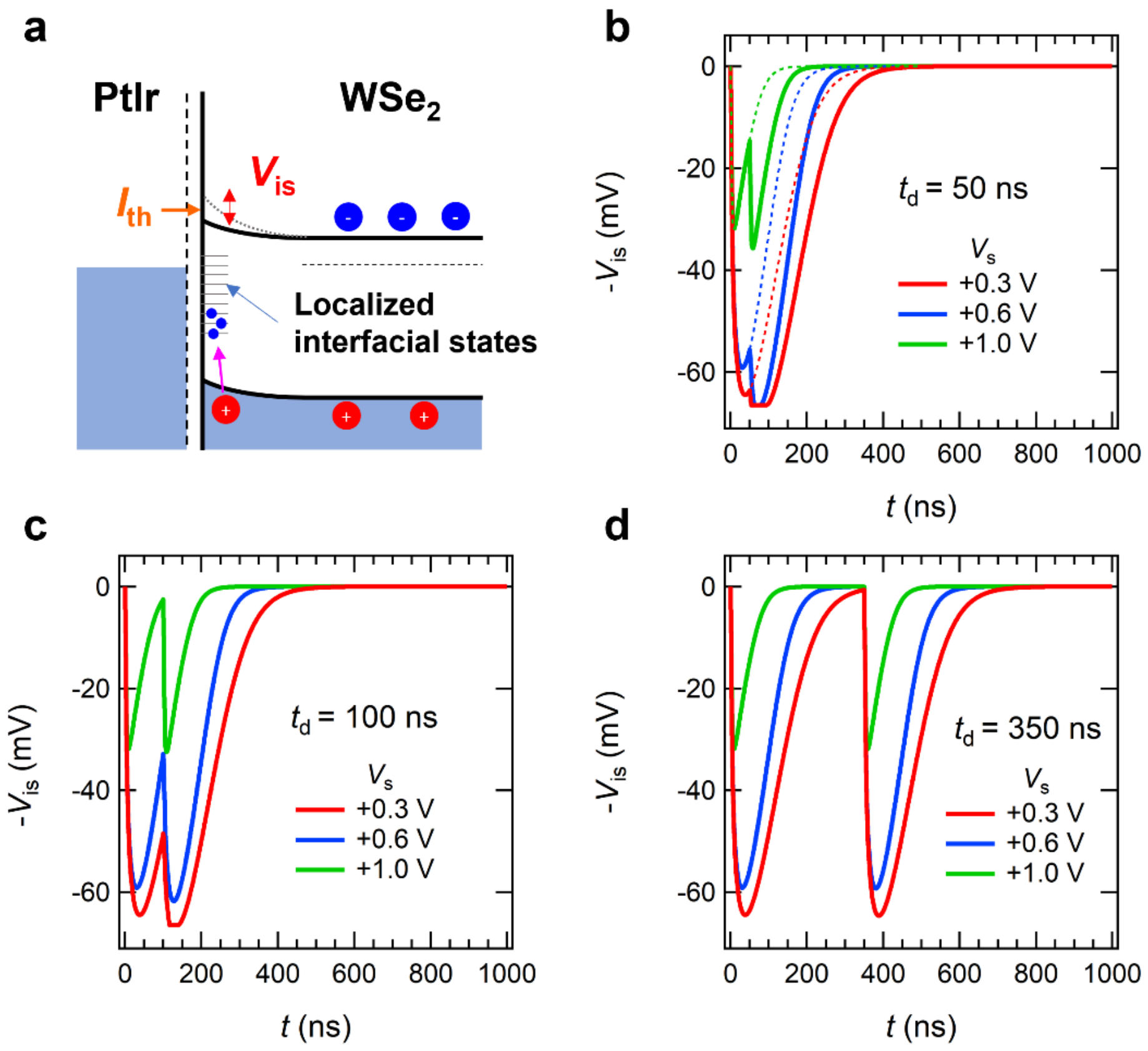

**Fig. S10: Simulated evolution of the photoinduced interfacial potential.**

(a) Schematic illustration of the photoinduced interface potential $V_{\mathrm{is}}$ generated by holes accumulated in localized interfacial states at the $WSe_2$/PtIr junction. The accumulated holes partially screen the built-in electric field, resulting in modulation of the effective Schottky barrier. (b–d) Simulated temporal evolution of the interface potential $V_{\mathrm{is}}(t)$ for representative pump–probe delay times of $t_{\mathrm{d}} = 50$, 100, and 350 ns, respectively. Red, blue, and green correspond to sample biases of $V_{\mathrm{s}} = +0.3$, +0.6, and +1.0 V, respectively. The solid curves represent the total interface potential produced by both the pump and probe pulses, whereas the dashed curves (shown only in panel b for clarity) represent the interface potential induced by the pump pulse alone. For intuitive visualization, the calculated interface potential is plotted as $-V_{\mathrm{is}}$.